\documentclass{article}
\usepackage{amsmath,amssymb}
\usepackage{graphicx}%
\usepackage{verbatim}
\def\stackunder#1#2{\mathrel{\mathop{#2}\limits_{#1}}}
\newcommand{\beq}[2]{\begin{equation}\label{#1}
#2 \end{equation}}
\newcommand{\beqdis}[2]{\begin{equation}\label{#1}
{\displaystyle #2} \end{equation}}
\newcommand{\dsp}{\displaystyle}

\newcommand{\krugskob}[1]{\left(#1\right)}
\newcommand{\kvadrskob}[1]{\left[#1\right]}
\newcommand{\figurskob}[1]{\left\{#1\right\}}
\newcommand{\Pg}{{\rm P}}

\newcommand{\Eps}{{\cal E}}
\newcommand{\Const}{\mathop{\rm Const}\nolimits}

\newcommand{\anti}[1]{\overline{#1} \,}

\begin{document}

\begin{center}
{\bf \Large Cosmological Evolution of Statistical System of Scalar Charged Particles}. \\[12pt]
Yurii Ignat'ev, Alexander Agathonov, Mikhail Mikhailov and Dmitry Ignatyev \\
N.I. Lobachevsky Institute of Mathematics and Mechanics, Kazan Federal University, \\ Kremleovskaya str., 35, Kazan, 420008, Russia
\end{center}

\begin{abstract}
In the paper we consider the macroscopic model of plasma of scalar charged particles, obtained by means of the statistical averaging of the microscopic equations of particle dynamics in a scalar field. On the basis of kinetic equations, obtained from averaging, and their strict integral consequences, a self-consistent set of equations is formulated which describes the self-gravitating plasma of scalar charged particles. It was obtained the corresponding closed cosmological model which also was numerically simulated for the case of one-component degenerated Fermi gas and two-component Boltzmann system. It was shown that results depend weakly on the choice of a statistical model. Two specific features of cosmological evolution of a statistical system of scalar charged particles were obtained with respect to cosmological evolution of the minimal interaction models: appearance of giant bursts of invariant cosmological acceleration $\Omega$ at the time interval $8\cdot10^3\div2\cdot10^4 t_{Pl}$ and strong heating ($3\div 8$ orders of magnitude) of a statistical system at the same times. The presence of such features can modify the quantum theory of generation of cosmological gravitational perturbations.

{\bf keywords}{\it physics of the early universe,	particle physics - cosmology connection, inflation, phantom scalar interaction}

\end{abstract}

\section{Introduction}
In the recent years there emerged a large number of articles devoted to scalar fields, which are necessary for an explanation of the secondary acceleration of the Universe. It is reasonable to single out models with a minimal interaction,  where the scalar field interacts with ``ordinary matter'' only gravitationally.  It is also reasonable to consider models with non-minimal interactions, where the scalar field interacts with ``ordinary matter'' by means of some coupling realized through a factor depending on the scalar field potential in the Lagrange matter function (potential connection) or on a factor depending on the scalar field derivatives (kinetic connection). These models are intended to describe, to some extent, at least the qualitative behavior of the Universe at large and small times. However, being essentially highly phenomenological, these models are unable to meet the aesthetic demands of the theorists and unlikely to become long-living. On the other hand, there exist theoretical models based on modified gravitation theories, which are able to describe the kinematics of cosmological extension.
These are, above all, A.Starobinsky's $f(R)$ gravitation models \cite{Starobinsky} as well as other theories such as the Poincare gauge theories of gravity\footnote{see e.g., A.V.Mickevich \cite{Minkevich}.}. By no means negating the possibility of modification of the Einstein theory, we nonetheless consider a cosmological model constructed with a fundamental scalar interaction since a scalar field being highly symmetric, can, on the one hand,  be a basis for elementary particle physics, and, on the other hand, realize the concept of physical vacuum. In this paper we consider phantom fields, i.e. fields possessing a negative ``kinetic'' energy.

In contrast to the above-mentioned nonminimal models of scalar interaction, we consider statistical systems of scalar charged particles in which some
sort of particles can directly interact with a scalar field by means of a  fundamental \textit{scalar charge}. On the other hand, a statistical system
possessing a non-zero scalar charge and being itself a source of a scalar field, can efficiently influence the scalar field, managing its behavior.
Such a scalar interaction was introduced in the general-relativistic theory in 1982 by one of the present authors \cite{Ignatev1,Ignatev2,Ignatev3,Ignatev4}
and a bit later by G.G.Ivanov \cite{Ivanov}. In particular, in %
\cite{Ignatev2,Ignatev3}, on the basis of kinetic theory, %
a self-consistent set of equations describing a statistical system of particles with scalar interaction was obtained. The article \cite{kuza} studied the group properties of equilibrium statistical configurations with a scalar interaction, and \cite{Ignat_Popov} established a tight relationship between the variable mass particle dynamics and the dynamics of a scalar charge particle. In the recent papers \cite{YuNewScalar1,YuNewScalar2,YuNewScalar3}\footnote{see also monographs \cite{Yubook1,Yubook2}}, the macroscopic theory of statistical systems with a scalar interaction was significantly improved and extended to the case of phantom scalar fields\footnote{see also \cite{Ignatev14_1} and the review \cite{Yu_stfi14}}.

Thereby, we consider a rigorous dynamic model of interaction of a scalar field with elementary particles, based on the Hamilton mi\-cro\-sco\-pic equations of motion and subsequent statistical averaging procedure. It turns out that the  interaction of a scalar field with particles can be introduced by a unique way; consequently, the ma\-cro\-sco\-pic equations of matter and scalar field
are obtained in a unique way by means of standard averaging procedures. Thereby a tight connection is established between micro and macro levels of a scalar interaction. Naturally enough, the model of a scalar interaction obtained in that way should have a more complex structure than the similar phenomenological models, but at the same time it also manifests more richer possibilities of the behavior\footnote{see e.g. \cite{YuMif}}.

\section{Dynamics of Particles with Scalar Interaction}
\subsection{Canonical Equations of Motion}
The canonical equations of relativistic particle motion relative to a pair of canonically conjugated dynamic variables $x^{i} $ (coordinates) and $P_{i} $ (generalized momentum) have the form (see e.g. \cite{Ignatev2})\footnote{Here and henceforth we use the universal system of units, $G=c=\hbar =1$}:

\begin{equation} \label{eqn:Eq1}
\frac{dx^{i} }{ds} =\frac{\partial H}{\partial P_{i} } ;\quad \quad \frac{dP_{i} }{ds} =-\frac{\partial H}{\partial x^{i} } ,
\end{equation}
where $H(x,P)$ is a relativistically invariant Hamilton function. Calculating the total derivative of the dynamic variables $\psi (x^{i} ,P_{k} )$, on account of (\ref{eqn:Eq1}) we obtain:

\begin{equation} \label{eqn:Eq2}
\frac{d\psi }{ds} =[H,\psi ],
\end{equation}
where the invariant Poisson brackets are introduced:
\begin{equation} \label{eqn:Eq3}
[H,\psi ]=\frac{\partial H}{\partial P_{i} } \frac{\partial \psi }{\partial x^{i} } -\frac{\partial H}{\partial x^{i} } \frac{\partial \psi }{\partial P_{i} } \; .
\end{equation}
As a result of (\ref{eqn:Eq3}), the Hamilton function is an integral of particle motion:
\begin{equation} \label{eqn:Eq4}
\frac{dH}{ds} =[H,H]=0,\Rightarrow H= \Const.
\end{equation}
The relation (\ref{eqn:Eq4}) could be named the normalization relation.
The invariant Hamilton function is determined ambiguously. Indeed, due to  (\ref{eqn:Eq3}), %
if $H(x,P)$ is a Hamilton function, any continuously dif\-fe\-ren\-ti\-able function $f(H)$ is also a Hamilton function.
The only possibility of introduction of an invariant Hamilton function squared in the generalized particle momentum in the presence of only gravitational and scalar fields is:
\begin{eqnarray} \label{eqn:Eq7}
H(x,P)=\frac{1}{2} \left[\psi(x)(P,P)-\varphi(x) \right],\\
\nonumber (P,P)\equiv g^{ik}P_iP_k,
\end{eqnarray}
and $\psi(x)$, $\varphi(x)$ are certain scalar functions.
Generally, a constant on right-hand side of (\ref{eqn:Eq4}) is chosen to be $1/2m^2$, where $m$ is the rest mass of a particle.
We will not do that, but will choose more convenient zero normalization of the Hamiltonian function \cite{Ignat_Popov,YuNewScalar1} since this constant can also be put into $\psi(x)$ function:
\begin{equation}\label{eqn:Eq7a}
H(x,P)=\frac{1}{2} \left[\psi(x)(P,P)-\varphi(x) \right]=0.
\end{equation}
In author's paper \cite{Ignatev1} relativistic - invariant Hamilton function of particles with the scalar charge $q$ being in a scalar field with potential $\varphi $ was introduced by relation:
\begin{equation} \label{eqn:Eq6}
H(x,P)=\sqrt{(P,P)} -q\varphi .
\end{equation}
Despite its compact form, this Hamilton function is not convenient due to existence of radical.
In other author's paper \cite{Ignatev1} relativistic - invariant function of Hamilton was introduced by relation:
\begin{equation} \label{eqn:Eq5}
H(x,P)=\frac{1}{2} m\left[\frac{(P,P)}{m+q\varphi } -q\varphi \right].
\end{equation}

Thus, from normalization relation (\ref{eqn:Eq7a}) we obtain:
\begin{equation}\label{eqn:Eq7b}
(P,P)=\frac{\varphi}{\psi},
\end{equation}
and from the first group of the canonical equations of motion (\ref{eqn:Eq1}) %
we obtain a relation between the generalized momentum and the particle velocity vector:
\begin{equation} \label{eqn:Eq10a}
u^{i} \equiv \frac{dx^{i} }{ds} =\psi P^{i} \Rightarrow P^{i} =\psi^{-1} u^{i} ,
\end{equation}
Substituting the last relation to the normalization condition (\ref{eqn:Eq7b}), we obtain:
$$(u,u)=\psi\varphi,$$
so that the normalization relation for the particle velocity vector
\begin{equation} \label{eqn:Eq11}
(u,u)=1.
\end{equation}
requires
$$\psi\varphi=1 \Rightarrow \psi=\varphi^{-1}.$$
Thereby, the particle's invariant Hamilton function could be defined by the single scalar function $\varphi(x)$. Taking into account the last relation, let us write down the Hamilton function in the final form:
\begin{equation}\label{eqn:Eq7 }
H(x,P)=\frac{1}{2} \left[\varphi^{-1}(x)(P,P)-\varphi(x) \right]=0,
\end{equation}
and from the canonical equations (\ref{eqn:Eq1}) we obtain a relation between the generalized momentum and the particle velocity vector:
\begin{equation}\label{eqn:Eq10}
P^i=\varphi \frac{dx^i}{ds}.
\end{equation}
From the definition(\ref{eqn:Eq7}) it follows that the ge\-ne\-ra\-li\-zed momentum vector is timelike:
\begin{equation}\label{eqn:Eq8}
(P,P)=\varphi^2.
\end{equation}
Let us notice a relation useful for future reasoning, which is a consequence of  (\ref{eqn:Eq3}), (\ref{eqn:Eq7}) and (\ref{eqn:Eq8}):
\begin{equation} \label{eqn:Eq9}
[H,P^{k} ]=\nabla ^{k} \varphi \equiv g^{ik} \partial _{i} \varphi;
\end{equation}
where $\nabla^i\equiv g^{ik}\nabla_k$ is the covariant derivative symbol.

\subsection{Equations of Motion in the Lagrangian Formulation}

From the second group of the canonical equations (\ref{eqn:Eq1}) we obtain the equations of motion in Lagrange formulation \cite{Yubook1}:

\begin{equation} \label{eqn:Eq12}
\frac{d^{2} x^{i} }{ds^{2} } +\Gamma _{jk}^{i} \frac{dx^{j} }{ds} \frac{dx^{k} }{ds} =\partial _{,k} \ln |\varphi|{\rm {\mathcal P}}^{ik} ,
\end{equation}
 where:

\begin{equation} \label{eqn:Eq13}
{\rm {\mathcal P}}^{ik} ={\rm {\mathcal P}}^{ki} =g^{ik} -u^{i} u^{k}
\end{equation}
is the tensor of an orthogonal projection onto the direction $u$ such that:

\begin{equation} \label{eqn:Eq14}
{\rm {\mathcal P}}^{ik} u_{k} \equiv 0;\quad {\rm {\mathcal P}}^{ik} g_{ik} \equiv 3.
\end{equation}
From these relations and the Lagrange equations (\ref{eqn:Eq12}) it follows a strict consequence of the velocity and acceleration vectors' orthogonality:

\begin{equation} \label{eqn:Eq15}
g_{ik} u^{i} \frac{du^{k} }{ds} \equiv 0.
\end{equation}
Let us note that the Lagrange equations of motion (\ref{eqn:Eq12}) are invariant relative to the sign of the scalar function $\varphi(x)$:
\begin{equation} \label{eqn:Eq16a}
\varphi(x)\rightarrow -\varphi(x).
\end{equation}
The Hamilton function (\ref{eqn:Eq7}) with its zero normalization $H\rightarrow -H$ is also invariant under the trans\-for\-ma\-ti\-on (\ref{eqn:Eq16a}). Therefore from Eqs. (\ref{eqn:Eq8}), \eqref{eqn:Eq10}, and the Lagrange equations \eqref{eqn:Eq12} it follows that the square of $\varphi $ scalar has a meaning of a squared \textit{ effective inertial mass of a particle, $m_{*} $, in a scalar field}:

\begin{equation} \label{eqn:Eq16}
|\varphi| =m_{*} .
\end{equation}
Let us note that the following action function %
formally coincides with the Lagrangian function of a relativistic particle with a rest mass $m_*$ in a gravitational field (see e.g. \cite{Land_Field}) corresponds to the cited choice of the Hamilton function:
\begin{equation} \label{eqn:Eq17}
S=\int  m_{*} ds.
\end{equation}

\subsection{Integrals of Motion}

Let us now find the existence conditions of a linear integral of the canonical equations of motion (\ref{eqn:Eq1}) which is related to the particle's total energy and momentum. To do so, we calculate the total derivative of the scalar product $(\xi ,P)$  with respect to the canonical parameter. Using the canonical equations of motion (\ref{eqn:Eq1}), the normalization condition (\ref{eqn:Eq8}), and the relation between the generalized and kinematic momenta (\ref{eqn:Eq10}), we find:

\begin{equation} \label{eqn:Eq18}
\frac{d(\xi ,P)}{ds} =\frac{1}{m_{*} } P^{i} P^{k} \mathop{L}\limits_{\xi } g_{ik} +\mathop{L}\limits_{\xi } m_{*} ,
\end{equation}
where $\mathop{L}\limits_{\xi } $ is a Lie derivative in the direction $\xi $ %
\footnote{see e.g. \cite{Petrov}.}, so that:
\begin{eqnarray}
\mathop{L}\limits_{\xi }a_i&=&\xi^ka_{i,k}+\xi^k_{~,i}a_k;\nonumber\\
\mathop{L}\limits_{\xi }a_{ij}&=&\xi^ka_{ij,k}+\xi^k_{~,i}a_{kj}+\xi^k_{~,j}a_{ik}.\nonumber
\end{eqnarray}
Assuming further

\begin{equation} \label{eqn:Eq19}
\frac{d(\xi ,P)}{ds} =0\Leftrightarrow (\xi ,P)={\kern 1pt} Const{\kern 1pt} ,
\end{equation}
taking into account the arbitrariness of the generalized momentum vector and its normalization condition, we obtain the conditions under which this equality holds:

\begin{equation} \label{eqn:Eq20}
\mathop{L}\limits_{\xi } g_{ik} =\rho g_{ik} \Rightarrow \rho =-\mathop{L}\limits_{\xi } \ln |m_{*} |.
\end{equation}
Substituting this result back to Eq. (\ref{eqn:Eq18}), we find the necessary and sufficient conditions for the existence of a linear integral of the canonical equations(see e.g. \cite{Yubook1}):

\begin{equation} \label{eqn:Eq21}
\mathop{L}\limits_{\xi } m_{*} g_{ik} =0.
\end{equation}
Thus \textit{ for the existence of the linear integral, it is necessary and sufficient that the conformally related space with the metrics $m_{*} g_{ik} $ allows a group of isometries with the Killing vector $\xi $}. Let us note that linear integrals (\ref{eqn:Eq19}) have the meaning of a total momentum (with a  spacelike vector $\xi $) or total energy (with a timelike vector $\xi $).

\subsection{The Choice of a Mass Function}

There emerges the question of choosing the function $m_{*} (\Phi )$. Let us highlight an important circumstance, not yet specifying this function. Consider the statistical fields $g_{ik} $ and $\Phi $, allowing a timelike Killing vector $\xi ^{i} =\delta _{4}^{i} $, so that the particle total energy $P_{4} $ is conserved. Furthermore, consider the reference frame in which  $g_{\alpha 4} =0$, so that coordinate $x^{4} $ coincides with the world time $t$. Then, from the relationship between the kinematic velocity $u^{i} $ and the particle total momentum vector $P_{i} $ (\ref{eqn:Eq10}) if follows:

\begin{equation} \label{eqn:Eq22}
P_{4} ds=m_{*} dt,
\end{equation}
where $P_{4} =E_{0} ={\kern 1pt} Const{\kern 1pt} >0$ is total energy of the charged particle. Therefore, if we want to conserve the same orientation of world and proper time (i.e. $dt/ds>0$), it is necessary to choose such a mass function which always remains nonnegative:

\begin{equation} \label{eqn:Eq23}
m_{*} >0.
\end{equation}
As can be seen, for example, in the Lagrange equations (\ref{eqn:Eq12}), it is convenient to choose the function so that:

\begin{equation} \label{eqn:Eq24}
m_{*} (\Phi )=|m_{*} (\Phi )|\ge 0.
\end{equation}
Further, on the one hand, in the absence of a scalar field or, more precisely, in a constant scalar field, the mass function should become the particle rest mass, $m\ge 0$.
On the other hand, the Lagrange equations (\ref{eqn:Eq12}), in case of a weak scalar field and small velocities, should become the classical equations of motion in a scalar field.  Hence, according to the correspondence principle, we should have:

\begin{equation} \label{eqn:Eq25}
m_{*} (0)=m;\quad \quad (m_{*} )_{,k} |_{\Phi =0} =q\Phi _{,k} ,
\end{equation}
where $q$ is a certain fundamental constant, \textit{the scalar charge of a particle}. Hence Eqs. (\ref{eqn:Eq25}) mean that at small values of the scalar potential $\Phi $ the function $m_{*} (\Phi )$ should have an expansion of the form:

\begin{equation} \label{eqn:Eq26}
m_{*} (\Phi ){\rm \simeq }m(1+\frac{q\Phi }{m} +..).
\end{equation}
To this condition corresponds the linear function $m_{*} (\Phi )=|m+q\Phi |$, which was used in the papers cited above.

Another, more radical approach, which is not contrary to the relations (\ref{eqn:Eq25}) and (\ref{eqn:Eq26}), may be suggested assuming that all the inertial mass of particles arises due to the interaction with a scalar field:
\begin{equation} \label{eqn:Eq27}
\varphi (\Phi )\equiv m_{*} =|q\Phi |.
\end{equation}
Then, the rest mass of a particle, $ m_ {0} $, is understood as its mass (\ref{eqn:Eq27}) at the present stage of the Universe evolution, related to the scalar potential $\Phi _{0} $: $m_{0} =m_{*} (\Phi (t_{0} ))$. This choice of  $\varphi (\Phi )$ function corresponds to the action function:
\begin{equation} \label{eqn:Eq28}
S=\int  |q\Phi |ds.
\end{equation}
The choice also meets the aesthetic criteria, since in this case, the Hamilton function (\ref{eqn:Eq27}) does not depend on the rest mass. On the other hand, it is clear that at a choice of $\varphi (\Phi )$ function in the form of (\ref{eqn:Eq27}), Lagrange equation (\ref{eqn:Eq12}) become symmetrical with respect to substitution $\Phi \to -\Phi $ and does not depend explicitly on the scalar charge under condition of $q\ne 0$:%%__

\begin{equation} \label{eqn:Eq29}
\frac{d^{2} x^{i} }{ds^{2} } +\Gamma _{jk}^{i} \frac{dx^{j} }{ds} \frac{dx^{k} }{ds} =(\ln |\Phi |)_{,k} {\rm {\mathcal P}}^{ik} ,
\end{equation}
In deriving Eq. (\ref{eqn:Eq29}) we took into account the differential identity: $d|y|={y\mathord{\left/ {\vphantom {y |y|}} \right. \kern-\nulldelimiterspace} |y|} dy$. However, due to normalization relation (\ref{eqn:Eq8}) for such a choice of the mass function, solutions still depend on a scalar charge in a form of dependence of energy on momentum $P_{4} (P^{2} )$.

\subsection{Example of One-Dimensional Motion}

Let us consider a following problem: suppose that there is a static scalar field in Minkowski space whose potential depends only on one coordinate, $x^{1} =x$, and let for simplicity $m_{*} =m+q\Phi =x$. Thus we have three Killing vectors, one is timelike and two others are spacelike:
 \def\stackunder#1#2{\mathrel{\mathop{#2}\limits_{#1}}}
$$\stackunder{1}{\xi}^i=\delta _{4}^{i};\; \stackunder{2}{\xi}^i=\delta _{2}^{i} ;\; \stackunder{3}{\xi}^i =\delta _{3}^{i} .$$
Accordingly, there are three linear integrals of motion:
\begin{eqnarray}\label{eqn:Eq30}
P_{2} =P_{2}^{0} ={\kern 1pt} Const{\kern 1pt} ;\; P_{3} =P_{3}^{0} ={\kern 1pt} Const{\kern 1pt} ;\;\\
\nonumber
P_{4} =P_{4}^{0} ={\kern 1pt} Const{\kern 1pt} .
\end{eqnarray}

Let for simplicity  $P_{2} =P_{3} =0$. Then, from the canonical equations of motions using normalization condition (\ref{eqn:Eq8}), we obtain the nontrivial equations
\begin{equation} \label{eqn:Eq31}
\frac{dx}{dt} =\mp \frac{\sqrt{P_{4}^{2} -x^{2} } }{P_{4} } .
\end{equation}
For simplicity, let us suppose $P_{4} =1$, $x(0)=1/2$. Then we select a negative sign on the right hand side in Eq. (\ref{eqn:Eq31}). The solution of this equation is:
\begin{equation} \label{eqn:Eq32}
x=\cos (t+\pi /3);
\end{equation}
and it describes harmonic oscillations by a world time clock $t$; while the generalized momentum, $P_{1} $ is also a harmonic function of a world time:

\begin{equation} \label{eqn:Eq33}
P_{1} =\sin (t+\pi /3).
\end{equation}
However, the coordinates of four-dimensional vector of kinematic particle velocity, $u^{i} $

$$u^{1} =\frac{dx}{ds} \equiv -P_{1} /\varphi =-\tan (t+\pi /3)$$
have second-order discontinuities at times $t=\pi /6+\pi k$, where $x=0$. This indicates either a break of connection between the coordinates and the proper time at the indicated world time, or the need to override the proper time for scalar charged particles. For the kinematic momentum of a particle, $p^{i} $, if we introduce it as

\begin{equation} \label{eqn:Eq34}
p^{i} =m_{*} \frac{dx^{i} }{ds} \equiv P^{i} ,
\end{equation}
this problem does not occur, as well as for the three-dimensional velocity $v^{\alpha } =u^{\alpha } /u^{4} $. Let us note that, actually, only those continuous quantities are physically measurable. Nevertheless, this example shows that it is necessary to perform accurate calculations for the scalar charged particles.

Further we will not specify normalization relation for effective mass, but we will just assume condition (\ref{eqn:Eq24}).

\subsection{Quantum Equations}
Let us note that using the standard procedure of obtaining quantum equations %
from the classical Hamilton function, it is necessary to change in the function:\footnote{Here we temporarily abandon the universal system of units, where $\hbar=1$.}
\begin{equation}\label{eqn:quant_trans}
P_i\rightarrow i\hbar \frac{\partial}{\partial x^i},
\end{equation}
As a result of a covariant generalization, the Hamilton operator is obtained from the Hamilton function:
\begin{equation}\label{eqn:Hamiltonian}
\hat{{\rm H}}=-m_*^{-1}(\hbar^2 g^{ik}\nabla_i\nabla_k + m^2_*).
\end{equation}
Thus for a free massive scalar field we can obtain the wave equations in form of the Klein-Gordon standard equations with the only difference that the
boson rest mass should be the effective mass \cite{Ignatev14_1}:
\begin{equation}\label{eqn:free_bozon}
(\square+m_*^2/\hbar^2)\Psi=0,
\end{equation}
and for free fermions we can obtained the co\-rres\-pon\-ding Dirac equations:
\begin{equation}\label{eqn:Dirac_eq}
(\hbar\gamma^i\nabla_i+m_*)\Psi=0,
\end{equation}
where $\gamma$ are the gamma matrices.

Let us note that from (\ref{eqn:free_bozon}), substituting $\Psi=\Phi$ and choosing the simplest mass function $m_*=|q\Phi|$, we immediately obtain the equation of a free scalar field with cubic nonlinearity:
\begin{equation}\label{eqn:Phi3}
\square\Phi+(q^2/\hbar^2)\Phi^3=0.
\end{equation}
Thus constant of self-action in a scalar field's cubic equation takes quite defined meaning:
$$\lambda=\frac{q^2}{\hbar^2}.$$

\section{Statistical Systems of Particles with Scalar Interaction}

Now let us describe a macroscopic scalar interaction. This description may be obtained by introducing invariant distribution functions of identical particles.

\subsection{The Distribution Function and Macroscopic Flux Densities}

Let $F(x,P)$ is an invariant particles distribution function in 8-dimensional
phase space, and let $\psi (x,P)$ be a certain tensor function of the dy\-namic variables $(x,P)$. According to \cite{Yubook1}, each tensor dy\-namic function can be associated with a macroscopic flux density:

\begin{eqnarray}\label{eqn:Eq35}
\Psi ^{i} (x)=\int\limits_{P(x)} F(x,P)\psi (x,P)\frac{\partial H}{\partial P_{i} } dP\equiv \\
\nonumber
 m_{*}^{-1} \int\limits_{P(x)} F(x,P)\psi (x,P)P^{i} dP,
\end{eqnarray}
Let us define, according to (\ref{eqn:Eq35}), the moments with respect to the distribution $F(x,P)$ \cite{Yubook1}:

\begin{eqnarray}\label{eqn:Eq36}
n^{i} (x)=\int _{P(x)} F(x,P)\frac{\partial H}{\partial P_{i} } dP\equiv \\
\nonumber
m_{*}^{-1} \int _{P(x)} F(x,P)P^{i} dP,
\end{eqnarray}
and the particle number flux density vector\footnote{A number vector according to J.L. Synge \cite{Sing,Sing1}.}:

\begin{equation} \label{eqn:Eq37}
n^{i} =nv^{i} ,
\end{equation}
where $v^{i} $ is a unit timelike vector of kinematic macroscopic particle velocity:
\begin{equation} \label{eqn:Eq38}
n=\sqrt{(n,n)} .
\end{equation}
Furthermore,
\begin{eqnarray} \label{eqn:Eq39}
T_{p}^{ik} (x)=\int _{P(x)} F(x,P)P^{i} \frac{\partial H}{\partial P_{k} } dP\equiv \\
\nonumber
m_{*}^{-1} \int _{P(x)} F(x,P)P^{i} P^{k} dP,
\end{eqnarray}
is a macroscopic energy-momentum tensor (EMT). Its trace can be calculated using the normalization condition (\ref{eqn:Eq8}):
\begin{equation} \label{eqn:Eq40}
T_{p} \equiv g_{ik} T_{p}^{ik} =m_{*} \int _{P(x)} F(x,P)dP.
\end{equation}
Next, the invariant volume element of 4-dimensional momentum space in the Eqs. %
(\ref{eqn:Eq36}), (\ref{eqn:Eq39}) in the adopted system of units is \cite{Ignatev2}:
\begin{equation} \label{eqn:Eq41}
dP=\frac{2S+1}{(2\pi )^{3} \sqrt{-g} } dP_{1} \wedge dP_{2} \wedge dP_{3} \wedge dP_{4} ,
\end{equation}
where $S$ is the particle spin. Thus the invariant 8-dimensional distribution function $F(x,P)$, which is singular on the mass surface (\ref{eqn:Eq8})
is connected with a nonsingular 7-dimensional distribution function $f(x,P)$ through a $\delta $-function by the relation \cite{Ignatev2}:
\begin{equation} \label{eqn:Eq42}
F(x,P)=f(x,P)\delta (H)=m_{*} \frac{\delta (P_{4} -P_{4}^{+} )}{P_{+}^{4} } ,
\end{equation}
where $P_{4}^{+} $ is a positive root of the normalization equation (\ref{eqn:Eq8});
$P_{+}^{4} =g^{4k} P_{k}^{+} $ is the value of the contravariant
momentum component corresponding to that root. In a local Lorentz frame of reference it is:
\begin{equation} \label{eqn:Eq43}
P_{+}^{4} =\sqrt{m_{*}^{2} +P^{2} } ,
\end{equation}
where $P^{2} =\sum _{\alpha =1}^{3} (P^{\alpha } )^{2} $ is the squared physical momentum.
Thus we obtain an invariant volume element of 3-dimensional momentum space:
\begin{equation} \label{eqn:Eq44}
dP_{+} =m_{*} \frac{2S+1}{(2\pi )^{3} \sqrt{-g} } \frac{dP_{1} \wedge dP_{2} \wedge dP_{3} }{P_{+}^{4} }
\equiv m_{*} dP_{0} ,
\end{equation}
where
\begin{equation} \label{eqn:Eq45}
dP_{0} =\sqrt{-g} \frac{2S+1}{(2\pi )^{3} } \frac{dP^{1} dP^{2} dP^{3} }{P_{4}^{+} }
\equiv \sqrt{-g} \frac{2S+1}{(2\pi )^{3} } \frac{d^{3} P}{P_{4}^{+} } .
\end{equation}
Then Eqs. (\ref{eqn:Eq36}), (\ref{eqn:Eq39}) and (\ref{eqn:Eq40}) take the following form (for simplicity we omit a summation in particle sorts):
\begin{eqnarray} \label{eqn:Eq46}
n^{i} (x)=\frac{2S+1}{(2\pi )^{3} } \int\limits_{P(x)} f(x,P)P^{i} \sqrt{-g} \; \frac{d^{3} P}{P_{4}^{+} } ;\\
\label{eqn:Eq47}
T_{p}^{ik} (x)=\frac{2S+1}{(2\pi )^{3} } \int\limits_{P(x)} f(x,P)P^{i} P^{k} \sqrt{-g} \; \frac{d^{3} P}{P_{4}^{+} } ;\\
\label{eqn:Eq48}
T_{p} =\frac{2S+1}{(2\pi )^{3} } m_{*}^{2} \int\limits_{P(x)} f(x,P)\sqrt{-g} \frac{d^{3} P}{P_{4}^{+} }.
\end{eqnarray}

\subsection{General Relativistic Kinetic Equations}
As a result of the local correspondence principle and the assumption that particle collisions are 4-pointed, in each particle interaction act, generalized momentum of the interacting particles system is conserved:
\begin{equation} \label{eqn:Eq49}
\sum _{I} P_{i} =\sum _{F} P'_{i} ,
\end{equation}
where summing is carried our in all initial, $P_{i} $, and final, $P'_{i} $, states. Let the following reactions proceed in the plasma:
\begin{equation} \label{eqn:Eq50}
\sum _{A=1}^{m} \nu _{A} a_{A} {\rm \rightleftarrows }\sum _{B=1}^{m'} \nu '_{B} a'_{B} ,
\end{equation}
where $a_{A} $ are particle symbols and $\nu _{A} $ are their numbers. Thus the generalized momenta
of initial and final states are equal to:
\begin{equation}\label{eqn:Eq51}
P_I=\sum\limits_{A=1}^m\sum\limits_\alpha^{\nu_A} P^\alpha_A,
\quad P_F=\sum\limits_{B=1}^{m'}\sum\limits_{\alpha'}^{\nu'_B} P'\
\!\!^{\alpha'}_B.
\end{equation}
The particle distribution functions are determined by the invariant kinetic equations \cite{Ignatev2}:
\begin{equation} \label{eqn:Eq52}
[H_{a} ,f_{a} ]=I_{a} (x,P_{a} ),
\end{equation}
where $I_{a} (x,P_{a} )$ is the collisions integral:
\begin{eqnarray} \label{eqn:Eq53}
I_{a} (x,P_{a} )=-\sum  \nu _{a} \int  '_{a} \delta ^{4} (P_{F} -P_{I} )\times\\
\nonumber
W_{IF} (Z_{IF} -Z_{FI} )\prod _{I,F} 'dP;
\end{eqnarray}
$$W_{FI} =(2\pi )^{4} |M_{IF} |^{2} 2^{-\sum  \nu _{A} +\sum  \nu '_{b} } $$
being the scattering matrix of the reaction channel (\ref{eqn:Eq50}) ($|M_{IF} |$ are the invariant scattering amplitudes);
$$Z_{IF} =\prod _{I} f(P_{A}^{\alpha } )\prod _{F} [1\pm f(P_{B}^{\alpha '} )];
$$ $$Z_{FI} =\prod _{I} [1\pm f(P_{A}^{\alpha } )]\prod _{F}
f(P_{B}^{\alpha '} ),$$ the plus sign corresponds to bosons and the minus to fermions (see details in \cite{Ignatev2,Ignatev3}).

\subsection{Transport Equations for Dynamic Quantities}
Strict con\-se\-qu\-ences of the general-relativistic kinetic equations (\ref{eqn:Eq52})
are the transport equations for the dynamic variables $\Psi _{a} (x,P_{a} )$ \cite{Ignatev3}:
\begin{eqnarray}\label{eqn:Eq54}
\nabla _{i} \sum _{a} \int _{P(x)}\!\!\! \!\Psi _{a} F_{a}
\frac{\partial H_{a} }{\partial P_{i} } dP_{a}\! -
\!\!\sum _{a} \int _{P(x)}\!\! F_{a} [H_{a} ,\Psi _{a} ]dP_{a} \nonumber\\
=-\!\!\!\sum _{by\; channels}\!\! \int \prod _{I,F} dP \left(\sum _{A=1}^{m} \nu _{A} \Psi _{A} -
\sum _{B=1}^{m'} \nu '_{B} \Psi '_{B} \right)\\
\nonumber
\hskip -24pt\times\delta ^{4} (P_{F} -P_{I} )(Z_{IF} W_{IF} -Z_{FI} W_{FI} ),
\end{eqnarray}
where summing carried out by all reaction channels (\ref{eqn:Eq50}).
Supposing $\Psi _{a} =P^{k} $ in (\ref{eqn:Eq54}),
where $g_{a} $ are certain fundamental charges which conserved in reactions (\ref{eqn:Eq50}). And on account of (\ref{eqn:Eq49}), (\ref{eqn:Eq51}) and (\ref{eqn:Eq54})
we obtain transport equation for particle flux densities
\begin{equation} \label{eqn:Eq55}
\nabla _{i} J_{G}^{i} =0,
\end{equation}
where
\begin{equation} \label{eqn:Eq56}
J_{G}^{i} =\sum _{a} \frac{2S+1}{(2\pi )^{3} } \; g_{a}
\int _{P(x)} f(x,P)P^{i} \sqrt{-g} \; \frac{d^{3} P}{P_{4}^{+} } .
\end{equation}
is the fundamental current density vector, corresponding to
charges $g_{a} $. In particular, the law of conservation
(\ref{eqn:Eq55}) is always valid for each type of particles
$b$ ($g_{a} =\delta _{a}^{b} $) under condition of elastic collisions between them. Supposing $\Psi _{a} =P^{k} $ in (\ref{eqn:Eq54}),
on account of (\ref{eqn:Eq9}), (\ref{eqn:Eq49}) and (\ref{eqn:Eq51}) energy-momentum transport equation for the plasma:
\begin{equation} \label{eqn:Eq57}
\nabla _{k} T_{p}^{ik} -\sigma \nabla ^{i} \Phi =0,
\end{equation}
where the \textit{scalar charge density}, $\sigma $, has been introduced in \cite{Ignatev4}:

\begin{equation} \label{eqn:Eq58}
\sigma =\frac{1}{2} \sum _{a} \frac{2S+1}{(2\pi )^{3} } \frac{dm_{*}^{2} }{d\Phi }
\int _{P(x)} f(x,P)\sqrt{-g} \; \frac{d^{3} P}{P_{4}^{+} } .
\end{equation}
In particular, choosing the mass function in the form of (\ref{eqn:Eq27}) an expression for the scalar charge density takes the form:
\begin{equation} \label{eqn:Eq59}
\sigma =\Phi \sum _{a} \frac{2S+1}{(2\pi )^{3} } q^{2}
\int _{P(x)} f(x,P)\sqrt{-g} \; \frac{d^{3} P}{P_{4}^{+} } .
\end{equation}
It should be noted that form of the EMT (\ref{eqn:Eq39}) or (\ref{eqn:Eq47}), as well as the that of scalar charge density (\ref{eqn:Eq58}) obtained for scalarly charged particles in \cite{Ignatev3}, under a given Hamilton function, is an unambiguous consequence of the suggestion of total momentum conservation in local particle collisions.
In particular, for a system of single-sort particles due to (\ref{eqn:Eq40}) and (\ref{eqn:Eq58}) we obtain the relation:
\begin{equation} \label{eqn:Eq60}
\sigma =\frac{d\ln m_{*} }{d\Phi } T_{p} .
\end{equation}
The choice of mass function in the form of (\ref{eqn:Eq27}) simplifies the expression and \textit{it becomes valid for multicomponent systems}:
\begin{equation} \label{eqn:Eq61}
\sigma =\frac{T_{p} }{\Phi } ,
\end{equation}

\section{Thermodynamic Equilibrium of the Plasma in a Gravitational Field}
\subsection{Local Equilibrium Distribution}
Entropy density of a statistical system is given by (see e.g. \cite{Landau_Stat})
\begin{equation}\label{eqn:1.5.h1}
\mathcal{ S}(x) = \sum\limits_a\int_{P(x)} dP_a  [\pm (1 \pm f_a) \ln(1 \pm f_a) - f_a \ln f_a]\,,
\end{equation}
so that:
$$\int \mathcal{S}(x)dV_\tau  =S(\tau)$$
is the total entropy of the system. Under conditions of thermodynamic equilibrium:
\begin{equation}\label{eqn:3.1.1}
\frac{d S}{d\tau} = 0\,.
\end{equation}
In \cite{Yubook1,Yubook2} it is shown that due to the unitarity of the $S$-matrix and the optical theorem, this can can be satisfied only under conditions:
\begin{equation}\label{eqn:3.1.2}
Z_{fi} - Z_{if} = 0
\end{equation}
in each channel reactions (\ref{eqn:Eq50}), as for T-invariant and for T-non-invariant reactions. Equation (\ref{eqn:3.1.2}) are analogous to the functional Boltzmann equations \cite{chern5} and have following solutions:
\begin{equation}\label{eqn:3.1.3}
F_a = {\rm e}^{- \phi_a}\krugskob{e^{- \phi_a} \mp 1}^{-1} \equiv \krugskob{1 \mp
{\rm e}^{\phi_a}}^{-1}\,,
\end{equation}
where $\phi_A$ are linear functions of pulses:
\beq{eqn:3.1.6}{ \phi_A(\Pg^{\alpha}_A) = - \lambda_A(x) + (\xi_A,
\Pg^{\alpha}_A)\,, }
$\xi^i (x)$ are vectors:
\beq{eqn:3.1.7}{ \xi^i_A (x) = \xi^i (x)\,, }
and $\lambda_A (x)$ are scalars satisfying the conditions:

\beq{eqn:3.1.8}{ \sum\limits_{A=1}^{N} \nu^k_A \lambda_A = 0\,, }
where $||\nu^k_A||$ is a matrix of integers \cite{Yubook1}.
The obvious condition of the closure of all reaction cycles
\begin{equation}\label{eqn:2.1.11}
{\rm rank}||\nu^k_A||<N
\end{equation}
is resulting in nontrivial solution of equations (\ref{eqn:3.1.8}).

{\it Conditions (\ref{eqn:3.1.7}), (\ref{eqn:3.1.8}) are the conditions of
local thermodynamic equilibrium (LTE); scalars $\lambda_A(x)$ are called the chemical potential of a statistical system.}

Substituting solutions (\ref{eqn:3.1.6}) in (\ref{eqn:3.1.3}) on account of (\ref{eqn:3.1.7}) we obtain {\it locally equilibrium distribution functions}:
\beq{eqn:3.1.9}{ f^0_a (x, \Pg_a) = \figurskob{\exp[ - \lambda_a +
(\xi, \Pg_a)] \mp 1}^{- 1}\,, }
whereas before, the upper sign corresponds to bosons, the lower to fermions.

The convergence of distribution moments (\ref{eqn:3.1.9}) implies vector $\xi^i(x)$
to be a timelike vector:
\beq{eqn:3.1.10}{ \xi^2 \equiv (\xi, \xi) > 0\,. }
Let us introduce a unit timelike field $v^i(x)$ using the $\xi^i(x)$:
\beqdis{eqn:3.1.11}{ v^i =
\frac{\xi^i}{\xi}\,; \quad (v,v) = 1\,, }
the local temperature $\theta (x)$ \cite{chern5}:
\beq{eqn:3.1.12}{ \theta (x) = \xi^{-1}}
and the chemical potentials, $\mu_a(x)$, in common normalization:
\beq{eqn:3.1.13}{ \mu_a(x) = \theta(x) \lambda_a(x)\,. }
Then distribution (\ref{eqn:3.1.9}) can be written as:
\beqdis{eqn:3.1.14}{ f^0_a(x,\Pg_a) = \figurskob{\exp
\kvadrskob{\frac{\dsp{- \mu_a + (v,\Pg_a)}}{\theta}} \mp 1}^{ -
1}\,. }

\subsection{Moments of the Equilibrium Distribution}
Let us compute the moments of the distribution (\ref{eqn:3.1.9}). It is convenient to go to the local Lorentz frame, where the time component is directed along the vector $v^i$. Further, we go over to a spherical coordinate system in the momentum space $P(X)$ and covariantly generalize obtained results. In the end, we obtain expressions for the particle number density components, $n^i_a(x)$, and the energy-momentum tensor for the $a$-th component of the plasma, $\stackunder{a}{T}^{ik}$, \cite{Ignatev3},
\cite{kuza}:
\begin{eqnarray}\label{eqn:3.1.17}
n^i_a(x) = n_a(x) v^i\,;\\
\label{eqn:3.1.18}
\stackunder{a}{T}^{ik}(x) = (\Eps_a + P_a) v^i v^k - P_a g^{ik}\,,
\end{eqnarray}
where:
\begin{eqnarray}\label{eqn:3.1.19}
n_a(x) =\frac{\rho}{2\pi^2} \int\limits_{0}^{\infty} P^2 d P\times\nonumber\\
\left\{\exp\kvadrskob{\frac{\dsp{- \mu_a + \sqrt{m^2_* + P^2}}}{\dsp
\theta}}  \mp 1\right\}^{- 1} \,; \\
\label{eqn:3.1.20}
\Eps_a(x) =
\frac{\rho}{2\pi^2} \int\limits_{0}^{\infty} \sqrt{m^2_* + P^2} P^2 d P\times\nonumber\\
\left\{\exp
\kvadrskob{\frac{\dsp{- \mu_a + \sqrt{m^2_* + P^2}}}{\dsp
\theta}} \mp 1\right\}^{- 1} ; \\
\label{eqn:3.1.21}
P_a(x) = \frac{\rho}{6\pi^2}
\int\limits_{0}^{\infty} \frac{P^4
d P}{\sqrt{m^2_* + P^2}}\times\nonumber\\
\left\{\exp \kvadrskob{\frac{\dsp{-
\mu_a + \sqrt{m^2_* + P^2}}}{\dsp \theta}} \mp 1\right\}^{- 1} \,.
\end{eqnarray}

Let us note that under the LTE conditions the chemical potential is equal to zero for massless particles which have zero fundamental charges.
This conclusion follows from the fact that the numbers $\nu^a_A$ of such particles involved in reactions (\ref{eqn:3.1.8}) can be completely arbitrary.
Then from the existence of particles-antiparticles annihilation follows well-known relation \cite{Landau_Stat}:
\beq{eqn:3.1.24}{\anti{\mu}_a = - \mu_a\,. }
Let us also note that a unit vector in a direction of particle %
flux density is called the {\it kinematic velocity of the medium}; a unit timelike %
 eigenvector of a particle EMT is called the {\it dynamic velocity of the medium} %
and an eigenvalue of the EMT, corresponding to this eigenvector, is called the %
{\it energy density of the medium} (see e.g. \cite{Sing}). %
Thus, under the LTE conditions the kinematic velocity of particles coincides %
with their dynamic velocity and it equals $V^i$. According to (\ref{eqn:3.1.18}) the particle energy density is equal to:
\beq{eqn:3.1.20a}{\stackunder{p}{\Eps}(x) = \sum\limits_{a}^{} \Eps_a\,,}
and the isotropic pressure of all the kinds of particles is
\beq{eqn:3.1.21a}{\stackunder{p}{P}(x) = \sum\limits_{a}^{} P_a\,, }
where $\Eps_a$ and $P_a$ are described by Eqs. (\ref{eqn:3.1.20}) and (\ref{eqn:3.1.21}).

\section{The Mathematical Model of a Self-Gravitating Plasma of Scalar Charged Particles}
The complete set of macroscopic equations, which describe a self-gravitating plasma of scalar charged particles ,first consists of the Einstein equations:
\begin{equation}\label{eqn:Einst_Scalar}
R^{ik}-\frac{1}{2}Rg^{ik}=8\pi (T^{ik}_p+T^{ik}_s),
\end{equation}
where $T^{ik}_p$ is the EMT of a statistical system defined above and $T^{ik}_s$ is the scalar field EMT:
\begin{equation}\label{eqn:Tik_s}
T_{s}^{ik} =\frac{\epsilon_1}{8\pi } \left[2\Phi ^{,i} \Phi ^{,k} -g^{ik} \Phi _{,j} \Phi ^{,j} +\epsilon_2 m_{s}^{2} g^{ik} \Phi ^{2} \right],
\end{equation}
where $\epsilon_2=1$ for a classical and $\epsilon_2=-1$ for a phantom scalar field; for a field with repulsion of equally charge particles  $\epsilon_1=1$, %
and for a field with attraction of equally charge particles $\varepsilon_1=-1$. Let us note that the scalar field EMT in the form (\ref{eqn:Tik_s}) is obtained from the Lagrangian \cite{YuNewScalar3}:
\begin{equation}\label{eqn:Ls}
L_s= \frac{\epsilon_1}{8\pi}(\Phi_i\Phi^{,i}-\epsilon_2 m^2_s\Phi^2).
\end{equation}
Secondly, on account of the transport equations for the particles' EMT, the requirement of the total EMT conservation gives two nontrivial relations: the scalar field with a source,
\begin{equation}\label{eqn:Eq_S}
\square\Phi+\epsilon_2m^2_s\Phi=-4\pi\epsilon_1\sigma,
\end{equation}
and the transport equation (\ref{eqn:Eq57}).

Thus, the strict macroscopic consequences of kinetic theory are transport equations, including the conservation law of a certain
vector current (\ref{eqn:Eq55}) which corresponds to the microscopic conservation law in reactions of a certain fundamental charge ${\rm e}$
(if such conservation law exists),
\begin{equation}\label{eqn:III.1}
\nabla_i\sum\limits_a {\rm e}_a n^i_a=0,
\end{equation}
as well as the statistical system energy-momentum conservation laws (\ref{eqn:Eq57}).
Let us ascertain what are the consequences of the conservation laws (\ref{eqn:Eq55}) -- (\ref{eqn:Eq57}) under conditions of the LTE.
From the normalization relation (\ref{eqn:3.1.11}) the well-known identity law follows:
\begin{equation}\label{eqn:III.6}
v^k_{~,i}v_k\equiv 0.
\end{equation}
Taking into account Eqs. (\ref{eqn:3.1.17}) -- (\ref{eqn:3.1.21}), the conservation laws (\ref{eqn:Eq57}) can be reduced to the form:
\begin{eqnarray}\label{eqn:III.7}
(\mathcal{E}_{pl}+\mathcal{P}_{pl})v^i_{~,k}v^k=(g^{ik}\!\!\!-v^iv^k)(\mathcal{P}_{pl,k}+\sigma\Phi_{~,k});\\
\label{eqn:III.7a}
\nabla_k(\mathcal{E}_{pl}+\mathcal{P}_{pl})v^k=(\mathcal{P}_{pl,k}+\sigma\Phi_{~,k})v^k,
\end{eqnarray}
and the conservation law of the fundamental charge ${\rm e}$ \ref{eqn:III.1}) becomes:
\begin{equation}\label{eqn:III.7b}
\nabla_k n_ev^k=0,\quad n_e\equiv \sum\limits_a {\rm e}_a n_a.
\end{equation}

\subsection{The Mathematical Model of a Scalar Charged Equilibrium Plasma}

Thus, formally for three macroscopic scalar functions $\mathcal{E}, \mathcal{P}, n_e$ and three %
independent components of velocity vector $u^i$ the macroscopic conservation laws give five independent %
equations (\ref{eqn:III.7}) -- (\ref{eqn:III.7b}), since equation (\ref{eqn:III.7}) depends on the other equations%
due to the identity (\ref{eqn:III.6}).
However not all specified macroscopic scalars
are functionally independent since they all are determined by the local equilibrium distribution functions(\ref{eqn:3.1.14}). %
Let us solve a series of chemical equilibrium conditions for a certain scalar potential and scale factor, and in the case of only one independent chemical potential at reduced mass surface equation. Then, four macroscopic scalar are determined by two thermodynamic scalar, which are certain chemical potential $\mu$ and local temperature $\theta$. Substituting
\begin{equation}\label{eqn:sht}
x=\sinh t,
\end{equation}
into equations (\ref{eqn:3.1.19}) -- (\ref{eqn:3.1.21}) let us reduce the expression for the macroscopic scalars to the form of \cite{Yu_stfi14,Ignat14_2}:
\begin{eqnarray} \label{eqn:na_LTE}
n_{a} =& \frac{2S+1}{2\pi ^{2} } m_{*}^{3} {\displaystyle\int _{0}^{\infty } \frac{{\rm sinh}^{2} x{\rm cosh}
xdx}{e^{-\gamma _{a} +\lambda _{*} {\rm cosh} x} \pm 1}} ;\\
\label{eqn:Ep_LTE}
{\rm {\mathcal E}}_{pl} =& \sum _{a} \frac{2S+1}{2\pi ^{2} } m_{*}^{4}
{\displaystyle\int _{0}^{\infty } \frac{{\rm sinh}^{2} x{\rm cosh}^{2} xdx}{e^{-\gamma _{a} +
\lambda _{*} {\rm cosh} x} \pm 1}}; \\
\label{eqn:Pp_LTE}
{\mathcal P}_{pl} = & \sum\limits _{a} \frac{2S+1}{6\pi ^{2} } m_{*}^{4} %
{\displaystyle\int_{0}^{\infty }\frac{{\rm sinh}^{4} xdx}{e^{-\gamma _{a} +\lambda _{*} {\rm cosh} x} \pm 1}}; \\
\label{eqn:Tp_LTE}
T_{p} = & \sum\limits_{a} \frac{2S+1}{2\pi ^{2} } m_{*}^{2} {\displaystyle\int _{0}^{\infty }
\frac{{\rm sinh}^{2} xdx}{e^{-\gamma _{a} +\lambda _{*} {\rm cosh} x} \pm 1}};\\
\label{eqn:s_LTE}
\sigma = & \sum\limits_{a} \frac{(2S+1)(m+q_a\Phi)^{3}}{2 \pi ^{2} } q
{\displaystyle\int _{0}^{\infty } \frac{{\rm sinh}^{2} xdx}{e^{-\gamma _{a} +\lambda _{*} {\rm cosh} z} \pm 1} },
\end{eqnarray}
where two dimensionless scalar functions are introduced:
\begin{equation}\label{eqn:lm}
\lambda _{*} =\frac{m_{*}}{\theta};\quad \gamma_a=\frac{\mu_a}{\theta}.
\end{equation}
Thus, the set of equations (\ref{eqn:III.7}) -- (\ref{eqn:III.7b}) is completely determined.

\subsection{The Cosmological Model}

Let us consider the above-formulated self-consistent mathematical model applied to a cosmological situation in a space-flat Friedman model:
$$ds^2=dt^2-a^2(t)(dx^2+dy^2+dz^2),$$
all thermodynamic functions depend only on time. It is easy to check that $u^i=\delta^i_4$
turns Eqs. (\ref{eqn:III.7}) into the identities and the set of equations (\ref{eqn:III.7a}) -- (\ref{eqn:III.7b}) %
is reduced to following two equations:
\begin{equation}\label{eqn:III.7a1}
\dot{\mathcal{E}}_{pl}+3\frac{\dot{a}}{a}(\mathcal{E}_{pl}+\mathcal{P}_{pl})=\sigma\dot{\Phi};
\end{equation}
\begin{equation}\label{eqn:III.7b1}
\dot{n}_e +3\frac{\dot{a}}{a}n_e=0.
\end{equation}
Thus, two differential equations remain with respect to two thermodynamic functions $\mu$ and $\theta$.

In this case the scalar field EMT also takes the form of a
perfect isotropic fluid EMT:
\begin{equation} \label{eqn:MET_s}
T_{s}^{ik} =({\rm {\mathcal E}}_s +{\rm {\mathcal P}}_{s} )v^{i} v^{k} -{\rm {\mathcal P}}_s g^{ik} ,
\end{equation}
where:
\begin{eqnarray}\label{eqn:Es}
{\rm {\mathcal E}}_s=\frac{\epsilon_1}{8\pi}(\dot\Phi^2+\varepsilon_2 m_s^2\Phi^2);\\
\label{eqn:Ps}
{\rm {\mathcal P}}_{s}=\frac{\epsilon_1}{8\pi}(\dot\Phi^2-
\varepsilon_2 m_s^2\Phi^2),
\end{eqnarray}
so that:
\begin{equation}\label{eqn:e+p}
{\rm {\mathcal E}}_s+{\rm {\mathcal P}}_{s}=\frac{\epsilon_1}{4\pi}\dot{\Phi}^2.
\end{equation}
The scalar field equation in the Friedman metrics takes the form:
\begin{equation}\label{eqn:Eq_S_t}
\ddot{\Phi}+3\frac{\dot{a}}{a}\Phi+\epsilon_2 m^2_s\Phi= -4\pi\epsilon_1\sigma(t).
\end{equation}
The last nontrivial Einstein equation must be added to these:
\begin{equation}\label{eqn:Einstein_a}
3\frac{\dot{a}^2}{a^2}=8\pi{\rm {\mathcal E}},
\end{equation}
where ${\rm {\mathcal E}}$ is the total energy density of the Fermi system and the scalar field. This set of equations describes a closed mathematical model of cosmological evolution of a degenerated Fermi system with scalar interaction (see e.g. \cite{Ignat14_2}).

\subsection{The Macroscopic Scalars for a Degenerated One-Component Fermi System}

It is interesting to investigate the dege\-ne\-rated Fermi system of scalar charged particles due to maximum simplicity of the mathematical model and the possibility of interpretation of such a system as a cosmological dark (cold) matter. We will not apply any constraints to the value of a particle charge. Thus, among others we will consider the situations when this value can be greater than one.

Under the total degeneracy conditions,
\begin{equation}\label{eqn:1}
\theta\to 0.
\end{equation}
the locally equilibrium distribution function of fermions (\ref{eqn:3.1.14}) takes the form of a step function \cite{Ignatev4}:
\begin{equation}\label{eqn:2}
f^0(x,P)=\chi_+(\mu-\sqrt{m_*^2+p^2}),
\end{equation}
where $\chi_+(x)$ is a step (Heaviside) function:
$$\chi_+(x)=\left\{%
\begin{array}{ll}
0,&x<0;\\
1,& x\geq 0.\\
\end{array}
\right.$$
Therefore integration of the macroscopic densities (\ref{eqn:na_LTE}) -- (\ref{eqn:s_LTE}) can be presented in elementary functions \cite{Ignatev4}:
\begin{eqnarray}\label{eqn:3}
n=\frac{1}{\pi^2}p_F^3;\\
\label{eqn:3a}
{\rm {\mathcal E}}_{pl} = \frac{m_*^4}{8\pi^2}
\bigl[\psi\sqrt{1+\psi^2}(1+2\psi^2)\\
\nonumber
-\ln(\psi+\sqrt{1+\psi^2})\bigr];\\
\label{eqn:3b}
{\mathcal P}_{pl} =\frac{m_*^4}{24\pi^2}
\bigl[\psi\sqrt{1+\psi^2}(2\psi^2-3)\\
\nonumber
+3\ln(\psi+\sqrt{1+\psi^2})\bigr];\\
\label{eqn:3c}
\sigma=q\frac{m_*^3}{2\pi^2}\left[\psi\sqrt{1+\psi^2}-
\ln(\psi+\sqrt{1+\psi^2})\right];
\end{eqnarray}
where dimensionless function $\psi$
\begin{equation}\label{eqn:psi}
\psi=p_F/m_*,
\end{equation}
is the Fermi momentum $p_F$ to effective fermion mass ratio.

At  $\mu\to0$ or $\theta\to0$ limit processing we obtain a set of equations on one function and a problem of contradictoriness of these equations
emerges at that this problem does not depend on presence of a scalar field. We will show that this problem is just apparent and
really there are no any contradictions in the set of equations (\ref{eqn:III.7a1}) -- (\ref{eqn:III.7b1}) emerging even in a case of the degenerated Fermi system.
Differentiating the energy density of the Fermi system (\ref{eqn:3a}) and taking into account the identity:
\begin{equation}\label{eqn:E_P_f}
\mathcal{E}_{pl}+\mathcal{P}_{pl}\equiv \frac{m^4_*}{3\pi^2}\psi^3\sqrt{1+\psi^2},
\end{equation}
let us reduce the energy conservation law for the Fermi system (\ref{eqn:III.7a1}) to the form of equation:
\begin{equation}\label{eqn:Eq_Pl}
\frac{d}{dt}\ln m_*\psi a=0.
\end{equation}
Hence, on account of $\psi$ (\ref{eqn:psi}) function definition we get:
\begin{equation}\label{eqn:ap}
ap_F={\rm Const}.
\end{equation}
This with an account of (\ref{eqn:3}) we obtain the fermion number conservation law:
\begin{equation}\label{eqn:na3}
a^3n={\rm Const}.
\end{equation}
Despite the apparent complexity of Eq. (\ref{eqn:III.7a1}) its solution can be easily found: from the Fermi system energy conservation law the fermion number conservation law is obtained.
Therefore, all we need for analyzing the cosmological evolution of a degenerate Fermi system is to integrate one Einstein equation and one field equation.

\subsection{The Macroscopic Scalars for the Boltzmann Two-Component Plasma}

Let us now consider a highly non-degenerate Boltzmann plasma;  first this means that reduced chemical potentials are small functions
\begin{equation} \label{eqn:gamma=0}
\gamma _{a} (t)\to 0
\end{equation}
and, secondly, statistical factors $\pm1$ in the denominators of (\ref{eqn:na_LTE})--(\ref{eqn:s_LTE}) can be neglected. Carrying out the integration in (\ref{eqn:na_LTE})--(\ref{eqn:s_LTE}), we obtain:
\begin{eqnarray}
\label{eqn:na_B}
n_{a} =\frac{2S+1}{\pi ^{2} } m_{*}^{3} \frac{\mathrm{K}_{2} (\lambda _{a} )}{\lambda _{a} } e^{\gamma _{a} } ; \\
\label{eqn:E_B}
\mathcal{E}_{pl} =\frac{1}{2\pi ^{2} } \sum _{a} (2S+1)m_{*}^{4} %
\left(\frac{\mathrm{K}_{3} (\lambda _{a} )}{\lambda _{a} } -\frac{\mathrm{K}_{2} (\lambda _{a} )}{\lambda _{a}^{2} } \right);  \\
\label{eqn:P_B}
\mathcal{P}_{pl} =\frac{1}{2\pi ^{2} } \sum _{a} (2S+1)m_{*}^{4} \frac{\mathrm{K}_{2} (\lambda _{a} )}{\lambda _{a}^{2} } ;\\
\label{eqn:s_B}
\sigma =\sum _{a} \frac{2S+1}{2\pi ^{2} } q_{a} (m+q_{a} \Phi )^{3} \frac{\mathrm{K}_{1}(\lambda _{a} )}{\lambda _{a}},
\end{eqnarray}
where
\begin{equation} \label{eqn:Kn}
\mathrm{K}_{n} (z)=\frac{\sqrt{\pi } z^{n} }{2^{n} \Gamma \left(n+\frac{1}{2} \right)} \int _{0}^{\infty }e^{-z cosh t} \sinh^{2n} tdt
\end{equation}
is Bessel function of imaginary order (see e.g. \cite{Lebed}) and $\Gamma (z)$ is gamma function.

For simplicity, in the paper we consider an ideal two-component plasma which consists only of scalar charged particles and antiparticles:
\begin{equation}\label{eqn:m+-}
m_{+} =|m+q\Phi |;\quad m_{-} =|m-q\Phi |
\end{equation}
such that particles have zero total chemical potential at the LTE conditions:
\begin{equation}\label{eqn:g+g=0}
\gamma +\bar{\gamma }=0\Rightarrow \bar{\gamma}=-\gamma.
\end{equation}
Obviously, when temperature increases the number of particle types is growing at the LTE.
But in this paper, for simplicity, we consider an idealized conditions, attempting to simplify the mathematical model. Now let us assume that some fundamental charge $\mathrm{e}$ is corresponded to these particles; then, according to (\ref{eqn:III.7b1}) in the case of a charge-symmetric plasma the charge conservation law should be satisfied:
\begin{eqnarray}\label{eqn:na3B}
m_{+}^{3} \frac{\mathrm{K}_{2} (\lambda _{+} )}{\lambda _{+} } e^{\gamma } - m_{-}^{3} %
\frac{\mathrm{K}_{2} (\lambda _{-} )}{\lambda _{-} } e^{-\gamma } =0\\
\nonumber
\Rightarrow e^{2\gamma } =\frac{m_{-}^{3} \lambda _{+} \mathrm{K}_{2} %
(\lambda _{-} )}{m_{+}^{3} \lambda _{-} \mathrm{K}_{2} (\lambda _{+} )}.
\end{eqnarray}
Thus, the chemical potential is determined completely for certain values of the plasma local temperature, $\theta$, and the scalar potential, $\Phi$.
In particular, under condition:
\begin{equation}\label{eqn:m+=-}
m_{-} =m_{+},
\end{equation}

which is realized only in two extreme cases:
\begin{equation}\label{eqn:g0_sym}
m_{-} =m_{+} =m_*\Rightarrow \left\{\begin{array}{llll}
1. & m & =0; & m_*=|q\Phi|\\
2. & q & =0; & m_*=m,
\end{array}\right.
\end{equation}
from (\ref{eqn:na3B}) we obtain the exact solution:
\begin{equation}\label{eqn:gamma=}
\gamma=0.
\end{equation}
Let us note that in these, and only these, cases (\ref{eqn:g0_sym}) scalar charged particles and antiparticles are symmetric; in other cases effective rest masses must be different at every moment of cosmological evolution. Thus, obviously, the second case describes {\it a minimal model} of scalar interaction.

In the paper, we will consider the case of {\it symmetric} Boltzmann plasma (\ref{eqn:g0_sym}), when particles and antiparticles have the same mass (\ref{eqn:m+=-}). In this case, it is convenient to choose required function as $\lambda_* (t)=\lambda _{-} =\lambda _{+}$, instead of the dimensional function $\theta (t)$. Expressions for the macroscopic scalars (\ref{eqn:E_B})--(\ref{eqn:s_B}) take the form of:
\begin{eqnarray} \label{eqn:E_sym}
{\mathcal E}_{pl} =\frac{2}{\pi ^{2} } m_{*}^{4} %
\left(\frac{\mathrm{K}_{3} (\lambda _{*} )}{\lambda _{*} } -%
\frac{\mathrm{K}_{2} (\lambda _{*} )}{\lambda _{*}^{2} } \right);\\
%%%%%%%%%%%%
\label{eqn:E+P_sym}
\mathcal{E}_{pl} +\mathcal{P}_{pl} =\frac{2}{\pi ^{2} } m_{*}^{4} \frac{\mathrm{K}_{3} (\lambda _{*} )}{\lambda _{*} } ; \\
\label{eqn:s_sym}
{\sigma =\frac{2q^{2} }{\pi ^{2} } \Phi \frac{\mathrm{K}_{1} (\lambda _{*} )}{\lambda _{*} } (3m^{2} +q^{2} \Phi ^{2} ).}
\end{eqnarray}

\section{The Numerical Simulation}
\subsection{The Cauchy Problem Definition}
On account of the particle number conservation law the problem of cosmological evolution of a statistical system with scalar charged particles %
is reduced to the solution of the set of two differential equations: %
the first-order equation (\ref{eqn:Einstein_a}) -- the Einstein equation and %
the second-order equation (\ref{eqn:Eq_S_t})-- the field equation. %
To set the Cauchy problem for the system (\ref{eqn:Einstein_a})--(\ref{eqn:Eq_S_t}) it %
is necessary to set the initial conditions for the values $a(t_{0} ),\Phi (t_{0} ),\dot{\Phi }(t_{0} ),\theta (t_{0} )$.  %
Let us hereinafter assume:
\begin{equation} \label{eqn:IC}
t_{0} =0;\quad a(0)=1;\quad \dot{\Phi}(0)=0.
\end{equation}
In the cosmological scenario corresponding to the initial conditions (\ref{eqn:IC})
at $t=0$ the scalar field kinetic energy is equal to zero and
the scalar field equation of state takes the form:
\begin{equation}\label{eqn:EP_s_IC}
{\mathcal P}_s(0)+{\rm {\mathcal E}}_{s}(0)=0.
\end{equation}
Let us note that the
choice of the initial time $t_0$ is arbitrary and the second condition in \eqref{eqn:EP_s_IC}  practically determine only %
the scale units and always can be realized. Therefore practically it is necessary to set only %
the initial conditions for the single sought function  $\Phi (t_{0})$ %
and to determine the constant in the relation (\ref{eqn:ap}). %
In the case of the Boltzmann system ``particle - antiparticle'' this constant is equal to zero, %
which leads to solution (\ref{eqn:na3B}) and in the case of a degenerate one-component Fermi systems %
this constant can be determined by means of the relationship (\ref{eqn:ap}) which on account of (\ref{eqn:IC}) takes the form: %%__
\begin{equation}\label{eqn:p0}
p_F(0)=p_0.
\end{equation}

However it is not very convenient for a numerical simulation to set the dimensional functions $p_0$ (or $\theta_0$ -- for the Boltzmann plasma) and $\Phi_0$ as the initial conditions. Instead we set two dimensionless scalar functions having the explicit physical meaning:
\begin{eqnarray}
\label{eqn:kp0}
\varkappa_{pl}(t)= \frac{\mathcal{P}_{pl}(t)}{\mathcal{E}_{pl}(t)}\Rightarrow & \varkappa^0_{pl}=\varkappa_{pl}(0); & \in [0,1/3); \\
\label{eqn:etas0}
\eta_S=\frac{\mathcal{E}_{S}(t)}{\mathcal{E}_{pl}(t)}\Rightarrow & \eta_S^0=\eta_S(0); & \in (-\infty,+\infty),
\end{eqnarray}
are an initial plasma's barotrope coefficient and an initial ratio of scalar field energy density to plasma energy density. %
Setting the relation $\varkappa^0_{pl}$ we can determine the initial Fermi momentum $p_0$ (or $\theta_0$) and setting $\eta_S^0$ we can determine $\Phi_0$.  Let us introduce the scalar functions $\varkappa(\mathcal{E})$ needed for the analysis :
\begin{equation} \label{eqn:kappa}
\mathcal{P}=\varkappa(\mathcal{E})\mathcal{E}\Rightarrow (\mathcal{P}_{s} +\mathcal{P}_{pl} )=\varkappa(\mathcal{E})(\mathcal{E}_{s} +\mathcal{E}_{pl} )
\end{equation}
is a total barotrope coefficient and  $\Omega (\mathcal{E})$:
\begin{equation} \label{eqn:Omega_kappa}
\Omega =\frac{a\ddot{a}}{\dot{a}^{2} } =-\frac{1}{2} (1+3\varkappa)
\end{equation}
is an invariant cosmological acceleration. In such setting the problem is determined by four independent initial conditions -- the nonvarying second and third conditions ((\ref{eqn:IC}) varying (\ref{eqn:kp0}) and (\ref{eqn:etas0}) ones and also by three parameters -- fundamental constants:  $m,q,m_{s}$. Thus there are five arbitrary constants in the problem.

\subsection{The Dimensionality of the Physical Values}

From the effective mass definition as well as the scalar field energy density definition
it follows the dimensionality of these values
in units of the Compton length
\begin{eqnarray}\label{eqn:diment}
[t]=l/c\to \ell; & [m]=[\mu]=\hbar/lc\to\ell^{-1}; &  \nonumber\\
\bigl[\mathcal{E}\bigr] \to \ell^{-4}; & \bigl[\Phi\bigr] =[m]=[\mu]\to \ell^{-1}; & [q]\to 1.
\end{eqnarray}
In ordinary units ($[m,l,t]$) the charge $q$ has the dimensionality of $m^{1/2} l^{3/2} t^{-1} $ and the scalar field potential has the one of $[\Phi ]=m^{1/2} l^{1/2} t^{-1} $. Thus in Planck units used in the article the value $q\Phi \sim 1$ corresponds to the effective mass of the scalar charged particles of the Planck mass order.

Further, since at numerical solution of the problem we deal with
the very large numbers, it is necessary to
scale the problem in advance. Let us introduce the dimensionless function instead of the scale factor:
\begin{equation}\label{eqn:L0}
\Lambda =\ln a(t);\quad \Lambda(0)=0,
\end{equation}
so that:
\begin{equation} \label{eqn:dotL}
\dot{\Lambda }=\frac{\dot{a}}{a} =H(t)
\end{equation}
is a Hubble constant, %
and invariant cosmological acceleration is expressed in terms of $\Lambda$ as follows:
\begin{equation}\label{eqn:Omega}
\Omega =1+\frac{\ddot{\Lambda }}{\dot{\Lambda }^{2} }
\equiv 1+\frac{\dot{H}}{H^{2}}
\end{equation}

Let us note that {\it the slow-roll approximation} commonly used in cosmological models (which we don't use in this paper):
\begin{equation}\label{eqn:slowly}
|\dot{H}|\ll H^2,
\end{equation}
and considering (\ref{eqn:Omega}) can be written as:
\begin{equation}\label{eqn:slowly_omega}
\frac{|\dot{H}|}{H^2}\equiv |\Omega-1|\ll 1\Rightarrow \Omega\rightarrow 1\Rightarrow \varkappa\rightarrow -1.
\end{equation}

\subsection{The Normal Set of Equations for the Fermi System}

The expression for $\psi$ function in the degenerate Fermi system model by means of introduced dimensionless variables can be written as following:
\begin{equation}\label{eqn:lps}
\psi=\frac{p_0}{m_*}{\rm e}^{-\Lambda}.
\end{equation}
For the numerical integration of the differential equations set let us bring them to the normal view assuming:
\begin{equation}\label{eqn:Z}
Z(t)=\dot{\Phi }
\end{equation}
and resolving the obtained system relative to the derivatives %
$\dot{\Lambda},\dot{\Phi }$ and  $\dot{Z}$, we obtain the normal set of equations:
\begin{eqnarray}
\label{eqn:norm1}
\dot{\Lambda}&=& \sqrt{\frac{8\pi}{3}}\sqrt{\mathcal{E}_{pl}+\mathcal{E}_{s}};\\
\label{eqn:norm2}
\dot{\Phi}&=&Z;\\
\label{eqn:norm3}
\dot{Z}&=&-3\dot{\Lambda}Z-\epsilon_2m^2_s\Phi-4\pi\epsilon_1\sigma,
\end{eqnarray}
where it is necessary to substitute the expressions for the Fermi system energy density (\ref{eqn:3a}) and scalar field (\ref{eqn:Es}) in the equation (\ref{eqn:norm1}) on account of the function $\psi$ (\ref{eqn:lps}) definition and equation (\ref{eqn:norm2}); the equation (\ref{eqn:norm3}) should have substituted in it the expression for $\dot{\Lambda}$ from the obtained equation (\ref{eqn:norm1}) and the expression for $\sigma$ obtained from the relation (\ref{eqn:3c}).

\subsection{The Normal Set of Equations for the Symmetric Boltzmann Plasma}

Let us obtain the normal set of differential equations for the symmetric Boltzmann plasma:
\begin{eqnarray}\label{eqn:EqLambdaB}
\dot{\Lambda }=&\sqrt{\frac{8\pi }{3} } \left\{\frac{2}{\pi ^{2} } m_{*}^{4} \left(\frac{\mathrm{K}_{3} (\lambda _{*} )}{\lambda _{*} } -\frac{\mathrm{K}_{2} (\lambda _{*} )}{\lambda _{*}^{2} } \right)\right.\\
\nonumber
&\left. +\frac{\varepsilon _{1} }{8\pi } (Z^{2} +\varepsilon _{2} m_{s}^{2} \Phi ^{2} )\right\}^{1/2};\\
\label{eqn:EqZB}\dot{Z}=&-Z\sqrt{24\pi } \left\{\frac{2}{\pi ^{2} } m_{*}^{4} \left(\frac{\mathrm{K}_{3} (\lambda _{*} )}{\lambda _{*} } -\frac{\mathrm{K}_{2} (\lambda _{*} )}{\lambda _{*}^{2} } \right)\right.\\
\nonumber
&\left. +\frac{\varepsilon _{1} }{8\pi } (Z^{2} +\varepsilon _{2} m_{s}^{2} \Phi ^{2} )\right\}^{1/2}-\varepsilon _{2} m_{s}^{2} \Phi \\
\nonumber
&-\varepsilon _{1} \frac{16q^{2} }{\pi } \Phi \frac{\mathrm{K}_{1} (\lambda _{*} )}{\lambda _{*} } (m^{2} +mq\Phi +q^{2} \Phi ^{2} ); \\
\label{eqn:EqF}\dot{\Phi }=&Z;\\                                                                                       \label{eqn:EqlambdaB}\nonumber
\dot{\lambda }_{*} =&{\displaystyle \frac{\lambda_*\dot{\Lambda}}{\displaystyle 1+\frac{\lambda_*\mathrm{K}_2(\lambda_*)}{3\mathrm{K}_3(\lambda_*)}}}+
\frac{q\dot{\Phi}}{m_*^4\left(\displaystyle 3\frac{\displaystyle \mathrm{K}_3(\lambda_*)}{\lambda_*^2}+ \displaystyle \frac{\displaystyle \mathrm{K}_2(\lambda_*)}{\lambda_*}\right)}\\
\nonumber
&\times\left[4(m+q\Phi)^3)\left(\displaystyle \frac{\displaystyle \mathrm{K}_{3} (\lambda _{*} )}{\lambda _{*} } -\displaystyle \frac{\displaystyle \mathrm{K}_{2} (\lambda _{*} )}{\lambda _{*}^{2}}\right)-\right.\\
&\left.q\Phi(3m^2+q^2\Phi^2)\displaystyle \frac{\displaystyle \mathrm{K}_1(\lambda_*)}{\lambda_*}
\right],
\end{eqnarray}
where we need to substitute the expression for $\dot{\Lambda }$ from Eq. (\ref{eqn:EqlambdaB}) to  Eq. (\ref{eqn:EqlambdaB}).

However, we must remember that considered symmetric problem can be implemented only in two cases (\ref{eqn:g0_sym}). Let us consider them in more detail. The second case describes a minimal interaction model.

\vskip 12pt
\textit{The minimal interaction model} $q=0$.  \hskip 12pt In this case, $m_{\pm } =m,\lambda _{\pm } =\lambda $; we must take into account this fact in Eq. (\ref{eqn:EqlambdaB}) and Eqs. (\ref{eqn:EqZB}), (\ref{eqn:EqlambdaB}) can be simplified as:
\begin{eqnarray}
\label{eqn:EqZBq=0}
\dot{Z}=-Z\sqrt{24\pi } \left\{\frac{2}{\pi ^{2} } m_{*}^{4} \left(\frac{\mathrm{K}_{3} (\lambda _{*} )}{\lambda _{*} } -\frac{\mathrm{K}_{2} (\lambda _{*} )}{\lambda _{*}^{2} } \right)\right.\\
\nonumber
\left.+\frac{\varepsilon _{1} }{8\pi } (Z^{2} +\varepsilon _{2} m_{s}^{2} \Phi ^{2} )\right\}^{1/2} -\varepsilon _{2} m_{s}^{2} \Phi ;\\
\label{eqn:EqlambdaBq=0}
\dot{\ln\lambda }=\dot{\Lambda }\left[1+\displaystyle\frac{1}{3} \frac{\lambda \mathrm{K}_{2} (\lambda )}{\mathrm{K}_{3} (\lambda )}\right]^{-1} .
\end{eqnarray}
Now let us consider Eq. (\ref{eqn:EqlambdaBq=0}) in the ultrarelativistic limit.
Taking into account the asymptotic behavior of the Bessel functions:
\begin{equation}\label{eqn:Knl0}
K_n(z)\stackunder{z\to 0}{\simeq}\frac{2^{n-1}\Gamma(n)}{z^n},
\end{equation}
we obtain:
\begin{equation}\label{eqn:lto0}
\ln C\lambda=\Lambda \Rightarrow \lambda\sim a(t)\Rightarrow \theta\sim a^{-1}(t)
\end{equation}
the law of cosmological evolution of a temperature in ultrarelativistic plasma.
\vskip 12pt
\textit{The model with zero rest mass} $m=0$. \hskip 12pt In this case $m_{\pm } =|q\Phi |$, $\lambda _{\pm } =\lambda _{*} =|q\Phi |/ \theta$ and equations (\ref{eqn:EqZB}), (\ref{eqn:EqlambdaB}) are modified as follows:
\begin{eqnarray}
\label{eqn:EqZBm=0}
\dot{Z}=-Z\sqrt{24\pi} \left\{\frac{2}{\pi ^{2}} m_{*}^{4} \left(\frac{\mathrm{K}_{3} (\lambda _{*} )}{\lambda _{*}} -\frac{\mathrm{K}_{2}(\lambda _{*} )}{\lambda _{*}^{2}} \right)\right.\\
\nonumber
\left.+\frac{\varepsilon_{1}}{8\pi} (Z^{2} +\varepsilon_{2}m_{s}^{2} \Phi ^{2})\right\}^{1/2}\\
\nonumber
-\varepsilon_{2}m_{s}^{2}\Phi-\varepsilon_{1}\frac{16q^{4}\Phi^3}{\pi} \frac{\mathrm{K}_{1} (\lambda_{*})}{\lambda_{*}};\\
\label{eqn:EqF}
\dot{\Phi}=Z;\\                                                                                       \label{eqn:EqlambdaBm=0}
\frac{d\ln\lambda_*}{dt} ={\displaystyle \frac{d\ln(a|\Phi|)}{dt}\frac{1}{\displaystyle 1+\frac{\lambda_*\mathrm{K}_2(\lambda_*)}{3\mathrm{K}_3(\lambda_*)}}}.
\end{eqnarray}
In the general case, Eq. (\ref{eqn:EqlambdaBm=0}) can also be solved in quadratures:
\begin{equation}\label{eqn:Gen_m=0}
\int\limits_0^{\lambda_*}\left[1+\frac{z\mathrm{K}_2(z)}{3\mathrm{K}_3(z)}\right]%
\frac{dz}{z}=a\Phi.
\end{equation}
Further, considering recurrence relations between the Bessel functions:
$$\frac{d}{dz}\mathrm{K}_n(z)z^n=-\mathrm{K}_{n-1}(z)z^n,$$
we can calculate the integral (\ref{eqn:Gen_m=0}) and obtain:
\begin{equation}\label{eqn:SolGen_m=0}
\mathrm{K}_3(\lambda_*)=(a\Phi-\Phi_0)^{-3}\mathrm{K}_3(\lambda^0_*)
\end{equation}
the algebraic relation between the functions $a(t)$, $\Phi(t)$, $\theta(t)$. In this case, %
like in the case of the degenerate Fermi systems, only two differential equations %
(three for normal system) remain. Let us note that from (\ref{eqn:SolGen_m=0}) in the %
ultrarelativistic limit we obtain asymptotically exact solution:
\begin{equation}\label{eqn:l_*_asypt}
\lambda_*\stackunder{\lambda_*\to0}{\approx}\lambda_0(a\Phi-\Phi_0)\Rightarrow \frac{|\Phi|}{\theta}\approx \frac{|\Phi_0|}{\theta_0}(a\Phi-\Phi_0),
\end{equation}
which for sufficiently large values of scale factor, gives again the ultrarelativistic law of the inverse temperature evolution (\ref{eqn:lto0}):
\begin{equation}\label{eqn:l*to0}
\lambda_*\sim a(t)|\Phi|\Rightarrow \theta(t)\sim a^{-1}(t).
\end{equation}

\section{The Numerical Simulation Results}

\subsection{Algorithms for the Numerical Simulation}

The numerical integration of the set of equations (\ref{eqn:norm1})--(\ref{eqn:norm3}) was carried out in the applied mathematics package ``Mathematica 9''. Since the set of differential equations reveals the signs of stiffness, there was used the numerical method with an automatic switching from the method ``stiff' to the precise Runge-Kutta method in form of
\begin{verbatim}Method ->{StiffnessSwitching,
Method ->{ExplicitRungeKutta,Automatic}},
AccuracyGoal -> 20, PrecisionGoal -> 5,
MaxSteps ->20000
\end{verbatim}
with a maximum number of steps is equal to 20000. Below there are shown the certain results of the simulation. For a convenient classification we will speak of the scalar field as a field of bosons, besides in the case of $m_s\ll 1$ we will speak of light bosons and in the case of $m_s\backsimeq 1$ we will speak of heavy bosons. The same is for fermions: $m\ll 1$ are light fermions, $m\backsimeq 1$ are heavy ones; also  $q\backsimeq 1$ are heavy fermions because of their potential energy. Next, in  the case of $\eta_S\ll 1$ we will speak of the fermion dominated system, in the case of  $\eta_s\gg 1$ we will speak of the boson dominated system. Everywhere in the presented plots it is  $m_s=0,001$ and the classification is carried out by the initial values of $\varkappa^0_{pl}$ and $\eta^0_S$.

Further, for plotting simulated results we will use author's function:
\begin{equation}\label{eqn:Lig}
\mathrm{Lig}(x)\equiv \mathrm{sgn}(x)\lg(1+|x|),
\end{equation}
which is required to represent results on a logarithmic scale when displayed function %
can change its sign. Introduced function $\mathrm{Lig}(x)$ is convenient due to its %
behavior, since for small argument values the function coincides with argument value, %
and for large argument values the function coincides with its decimal logarithm taken %
with the sign of the argument:
$$\mathrm{Lig}(x) \approx \left\{%
\begin{array}{ll}%
x, &  |x|\to 0;\\
\mathrm{sgn}(x)\ln|x|, & |x|\to\infty.
\end{array}
\right.
$$
It is easy to show that
$$\frac{d}{dx}\mathrm{Lig}(x)\geq 0,$$
and that provides a continuously differentiability of the function $\mathrm{Lig}(x)$, %
and this, in turn, provides bijective mapping of (\ref{eqn:Lig}). Inverse transformation %
can be done by expression:
$$%
x=\left\{\begin{array}{ll}
10^{\mathrm{Lig}(x)}-1, & x\geq 0;\\
1-10^{-\mathrm{Lig}(x)}, & x<0.
\end{array}
\right. $$

Further, the main results of numerical simulation will be presented in graphical form, grouped in pairs (with some exceptions), where the first plot corresponds to the case of a degenerate Fermi system and the second plot corresponds to a two-components Boltzmann plasma. The results for each pair were obtained at the same values of the fundamental constants $m,q,m_s$ and the same initial conditions $\varkappa^0_{pl},\eta^0_S$ (\ref{eqn:kp0}) -- (\ref{eqn:etas0}).

\subsection{The Ultrarelativistic particles $\varkappa^0_{pl}=1/3; m=0$,  %
The Phantom Scalar Field, $m_s=0.001$, The Scalar Dominated System: $\eta^0_S=100$}
The following approximate initial values of the Fermi momentum $p_F^0$ and the scalar potential $\Phi_0$ %
in the case of Fermi system, correspond to the given initial parameters and constants:$\Phi_0$, --
$[q,p^0_F,\Phi_0]$:\\ $[0.001,0.0112, -0.99]$, $[0.1,0.0035,-0.100]$, \\$[1,0.0011, -0.0100]$, $[10,0.00035, -0.00100].$\\
Initial values and parameters for the Boltzmann plasma are %
$[q,\lambda_0,\Phi_0]$ with following approximate values: \\
$[0.001,0.141, 0.361]$, $[0.1, 0.141,0.0000361]$, \\$[1, 0.141, 3.61\cdot10^{-7}]$, %
$[10, 0.141421 0.141,3.61\cdot10^{-9}]$.

For this case we will give sufficiently detailed results of the numerical integration. Further, we will not detail results, since they exhibit a remarkable stability under the model parameters changing. We present the results of numerical simulation for specified above cases.
\begin{figure}[tb]
\includegraphics[width=\columnwidth]{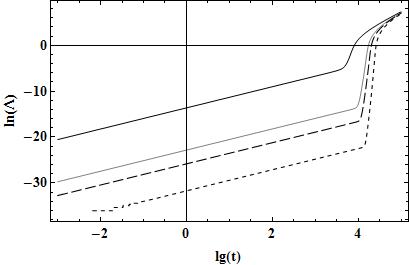}
\caption{The dependency of the scale function $\lg\Lambda(t)$
evolution on the fermion scalar charge value. Top to bottom: $q=0.001;0.1;1;10$. Lines for $q=0.1$ and $q=1$ are merged.}
%\label{fig:?}
\end{figure}
\begin{figure}[tb]
\includegraphics[width=\columnwidth]{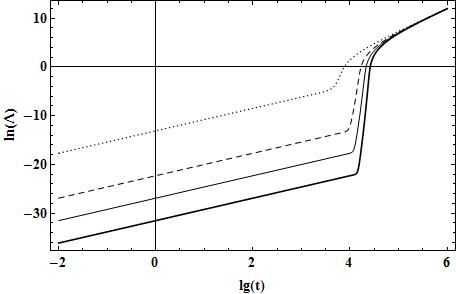}
\caption{The dependency of the scale function $\lg\Lambda(t)$
evolution on the scalar charge value for the Boltzmann plasma. Top to bottom: $q=0.001;0.1;1;10$.}
%\label{fig:?}
\end{figure}
On plots 3 and 4 it is shown the evolution of the scalar potential derivative of a  %
phantom field which is responsible for the negative kinetic energy; and on plots 5 %
and 6 it is shown the dependency of the invariant cosmological acceleration
$\Omega$ on the scalar charge value (more precisely, the function $\mathrm{Lig}\Omega$).
\begin{figure}%[tb]
\includegraphics[width=\columnwidth]{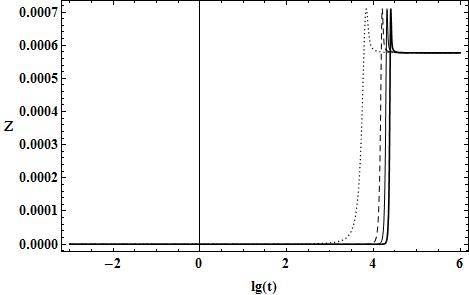}
\caption{The dependency of the potential derivative %
$Z=\dot{\Phi}$ evolution on the fermion scalar charge value. Left to right: $q=0.001;0.1;1;10$.}
%\label{fig:?}
\end{figure}
\begin{figure}%[tb]
\includegraphics[width=\columnwidth]{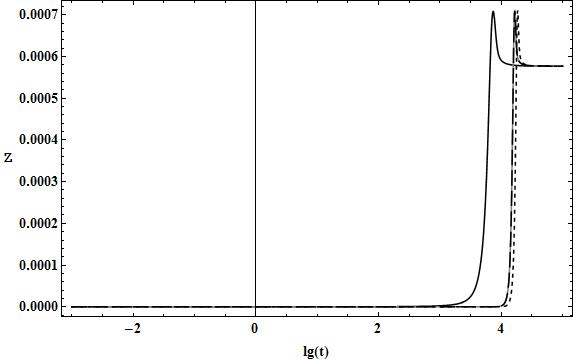}
\caption{The dependency of the invariant cosmological acceleration %
$\Omega$ evolution on the scalar charge value for the Boltzmann plasma. Bottom to top: $q=0.001;0.1;1;10$.}
%\label{fig:?}
\end{figure}

The presence of ``phantom stalagmites'' is a characteristic of the acceleration burst
at times of order $3\div 5\cdot10^3\ t_{pl}$. Plots of $\varkappa(t)$, on the contrary, %
contain ``phantom stalagmites'' at the same evolution times. It is necessary to notice %
that given phantom emissions are not the results of numerical calculations errors. %
This fact had been checked repeatedly with different models and at different accuracy %
of calculations (Fig. 5--8).
\begin{figure}%[tb]
\includegraphics[width=\columnwidth]{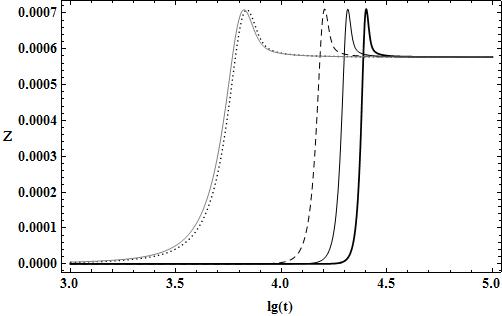}
\caption{The detailed structure of the phantom %
burst $Z$ depending on the value of particle's scalar charge for the Fermi plasma. Left to right: $q=0.001;0.1;1;10$.}
%\label{fig:?}
\end{figure}
\begin{figure}%[tb]
\includegraphics[width=\columnwidth]{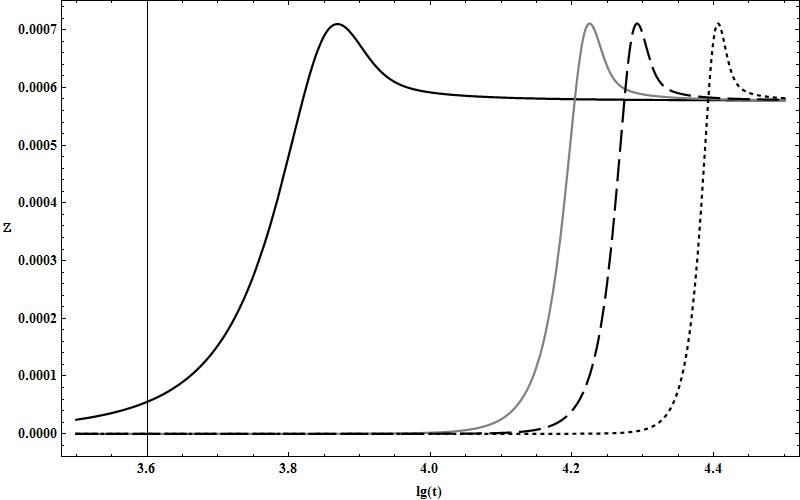}
\caption{The detailed structure of the phantom burst $Z$ depending on the value of particle's scalar charge for the Boltzmann plasma. Left to right: $q=0 (m=0.001);q=0.001;0.1;1;10$.}
%\label{fig:?}
\end{figure}

\begin{figure}%[tb]
\includegraphics[width=\columnwidth]{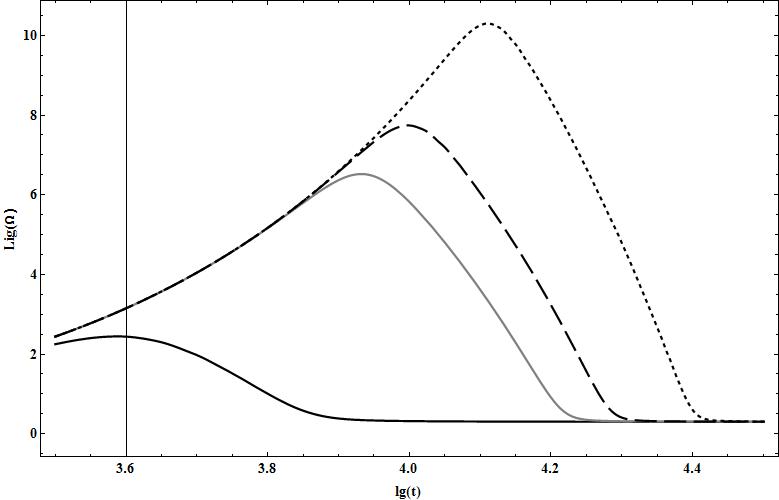}
\caption{The detailed structure of the phantom %
burst of the invariant cosmological acceleration $\Omega$ depending on the value of particle's scalar charge for the degenerate Fermi system. Left to right: $q=0.001;0.1;1;10$.}
%\label{fig:?}
\end{figure}
\begin{figure}%[tb]
\includegraphics[width=\columnwidth]{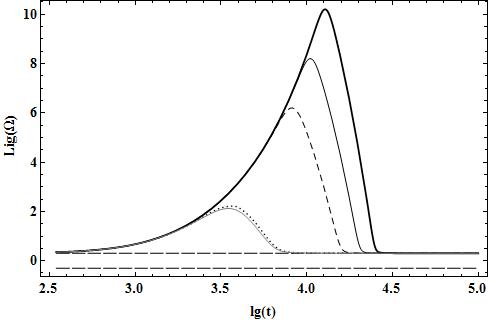}
\caption{The detailed structure of the phantom burst of the invariant cosmological acceleration $\Omega$ depending on the value of particle's scalar charge for the Boltzmann gas. Left to right: $q=0 (m=0.001); q=0.001;0.1;1;10$, large-scale structure of the burst is shown on plot 10.}
%\label{fig:?}
\end{figure}

Specified phantom stalagmites correspond to phantom stalactites on the plots of scalar potential. %
The scalar potential derivative reaches a constant value, which is almost independent on the scalar particle charge value. Thus, the scalar field potential value after the burst goes to the asymptote:
\begin{equation}\label{eqn:Phi_as}
\Phi\approx\Phi(t_i)+0.0006(t-t_i),
\end{equation}
where $t_i\sim 3\cdot10^4t_{Pl}$ is the end burst time. On plots 9, 10 the comparison of %
the cosmological acceleration evolutions is shown; horizontal dashed lines correspond to %
the values of $\Omega=-1$ (bottom) and $\Omega=+1$ (top).
\begin{figure}%[tb]
\includegraphics[width=\columnwidth]{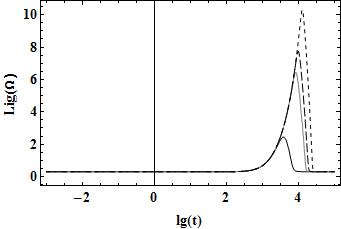}
\caption{The dependency of the invariant cosmological acceleration %
$\Omega$ evolution on the value of the fermion scalar charge. Bottom to top: %
$q=0.001;0.1;1;10$; lines for $q=0.1$ and $q=1$ are merged.}
%\label{fig:?}
\end{figure}
\begin{figure}%[tb]
\includegraphics[width=\columnwidth]{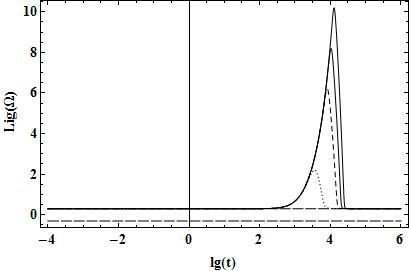}
\caption{The dependency of the invariant cosmological acceleration %
$\Omega$ evolution on the value of the scalar charge for the Boltzmann plasma. Bottom to top: $q=0.001;0.1;1;10$.}
%\label{fig:?}
\end{figure}
Let us note that the appearance of sharp phantom bursts of a super acceleration %
(an acceleration of order $10^2\div10^{10}$!) is a characteristic feature of our %
non-minimal interaction model; they occur at times of order %
$8\cdot10^3\div2\cdot10^4 t_{Pl}$ with a simultaneous increasing of the negative %
kinetic energy of a phantom field at the scalar charge value $q\gtrsim 0.1$. %
The model enters the standard inflationary expansion $\Omega\to 1$ after acceleration bursts. %
It is obvious that the existence of such bursts strongly violates the slow-roll %
approximation (\ref{eqn:slowly}), which, in turn, is an essential part of the overwhelming %
majority of cosmological models. %
Let us also note that these bursts occur at times of less than $10^5\div10^6 t_{Pl}$, %
and according to common theories (see e.g. \cite{Star}), the {\it flat} spectrum of gravitational perturbations is formed at that times. %
The presence of {\it very significant} violations of the slow-roll approximation can probably affect the final spectrum shape of gravitational perturbations and leave characteristic ``serifs'' on it. We also note that this behavior is found even for scalar-dominated system.

On plots 11, 12 it is shown the dependency of the total energy density $\mathcal{E}$ evolution on the value of the particle scalar charge for the Fermi and Boltzmann systems.

From these plots one can see that, first of all, the plot of the total energy density %
goes to the common asymptote at variables $(\lg t,\lg\mathcal{E})$, which corresponds %
to the total energy density power law. Secondly, the ratio of scalar field energy %
density to the Fermi system energy density increases rapidly reaching huge values of %
$\eta_S\sim 10^{300}$ at $t\sim 10^5$! The given example shows that at large times %
$t>10^4$ the model practically does not depend on the value of charge and and behaves %
like a minimal model. However at $t<10^4$ the model behavior significantly depends on %
the scalar charge value.
\begin{figure}%[tb]
\includegraphics[width=\columnwidth]{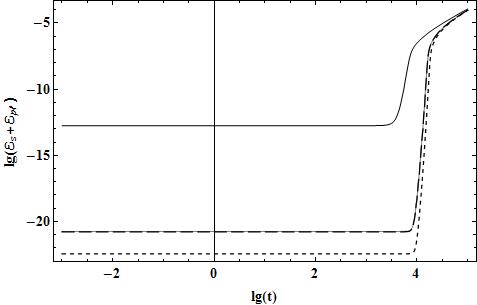}
\caption{The dependency of decimal logarithm of the invariant total energy density $\lg\mathcal{E}$
evolution on the value of the fermion scalar charge. Top to bottom: $q=0;001;0.1;1;10$.}
\label{fig:lgEsEp1}
\end{figure}
\begin{figure}%[tb]
\includegraphics[width=\columnwidth]{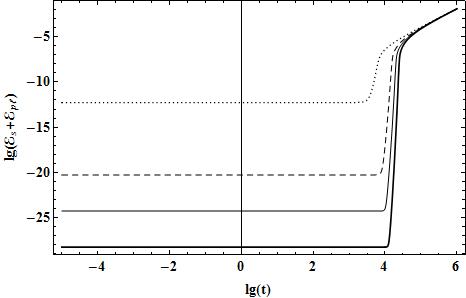}
\caption{The dependency of decimal logarithm of the invariant total energy density $\lg\mathcal{E}$
evolution on the scalar charge value for the Boltzmann gas. Top to bottom: $q=0;0.001;0.1;1;10$.}
\label{fig:lgEsEp1B}
\end{figure}

Further, on plots 13 and 14 it is shown the cosmological evolution of a temperature %
and scalar charge density for the Boltzmann system. On plot 13, the best marked plasma %
temperature bursts are shown at the same time as the phantom bursts of  the cosmological %
acceleration. The magnitude of the burst temperature increases with the particle %
scalar charge value. Bursts of the temperature, as well as bursts of acceleration, %
also appear when $q\geq 0.1$. As can be seen from the plots, the plasma temperature at %
times $8\cdot10^3\div2\cdot10^4 t_{Pl}$ can be increased by $3\div 8$ orders of %
magnitude. Let us note that there is a sharp decreasing of the scalar charge density %
after phantom burst of the acceleration (Fig. 14).
\begin{figure}%[tb]
\includegraphics[width=\columnwidth]{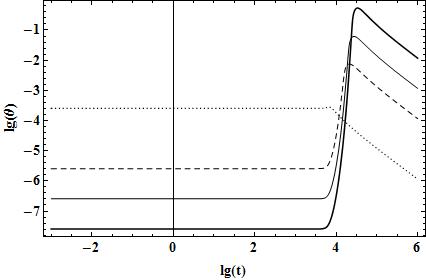}
\caption{The dependency of decimal logarithm of the Boltzmann gas temperature $\theta$ evolution on the value of the scalar particle charge. Bottom to top: $q=0;0.001;0.1;1;10$.}
\label{fig:lgtheta1}
\end{figure}
\begin{figure}%[tb]
\includegraphics[width=\columnwidth]{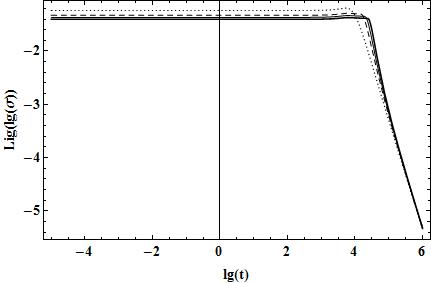}
\caption{The dependency of the scalar charge density function $\mathrm{Lig}\lg(\sigma)$ evolution on the scalar charge value for the Boltzmann gas. Bottom to top: $q=0;0.001;0.1;1;10$.}
\label{fig:lgsigmaB2}
\end{figure}

\subsection{The Ultrarelativistic Fermions $\varkappa^0_{pl}=1/3; m=0$, $m_s=0.001$, %
The Phantom Scalar Field, The Plasma Dominated System: $\eta^0_S=0.01$}
The following approximate initial values of the Fermi momentum $p_F^0$ and the scalar potential $\Phi_0$ in the case of Fermi system, correspond to the set-up initial parameters and constants:
$[[q,p^0_F,\Phi_0]$:\\ $[[0.001,0.000512,0.0021]$, $[0.1,5.15\cdot10^{-6}, 2.11\cdot10^{-7}]$, %
$[1, 5.05\cdot10^{-6}, 2.04\cdot10^{-7}]$, $[10, 1.92\cdot10^{-6}, 3.00\cdot10^{-8}]].$\\

Initial values and parameters for the Boltzmann plasma are %
$[q,\lambda_0,\Phi_0]$ with following approximate values: \\
$[0.001,0.141, 36.1]$, $[0.1,0.141,0.0036]$, \\$[1, 0.141, 1, 0.000036]$, %
$[10, 0.141,3.61199*10\cdot^{-7}]$.

In this case the following results were obtained (Fig. 15 -- 20)
\begin{figure}%[tb]
\includegraphics[width=\columnwidth]{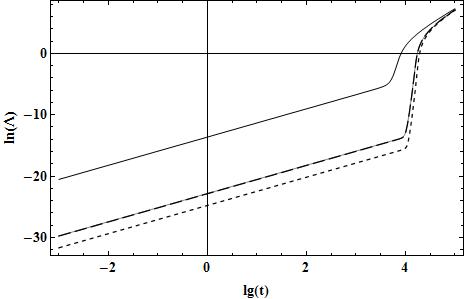}
\caption{The dependency of the scale function $\lg\Lambda(t)$ evolution on the fermion scalar charge value. Top to bottom: $q=0.001;0.1;1;10$. Lines for $q=0.1$ and $q=1$ are merged.}
%\label{fig:?}
\end{figure}
\begin{figure}%[tb]
\includegraphics[width=\columnwidth]{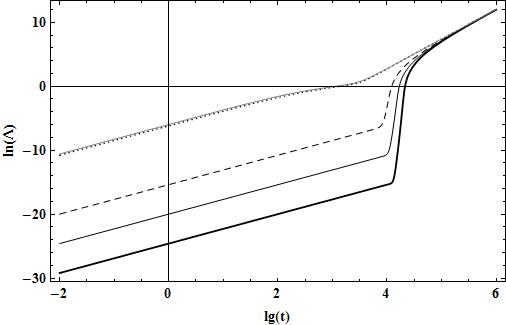}
\caption{The dependency of the scale function $\lg\Lambda(t)$
evolution on the scalar charge value for the Boltzmann plasma. Top to bottom: $q=0.001;0.1;1;10$.}
%\label{fig:?}
\end{figure}
Additionally, grey line on the plot (merged with dotted line) shows the evolution of the scale function for a minimal model with $m=0.04$.

We can see that the appearance of phantom bursts of the acceleration is very stable before entering the standard inflationary regime at times of order $8\cdot10^3\div2\cdot10^4 t_{Pl}$ in a system of scalar charged particles. Let us note that despite of the described case, here the initial ratio of scalar field energy to plasma energy is less than 4 orders of magnitude. The temperature burst is also stable when $q\geq 0.1$ at this time interval. Its magnitude also decreases with the increasing of the initial power of plasma energy dominance.
\begin{figure}%[tb]
\includegraphics[width=\columnwidth]{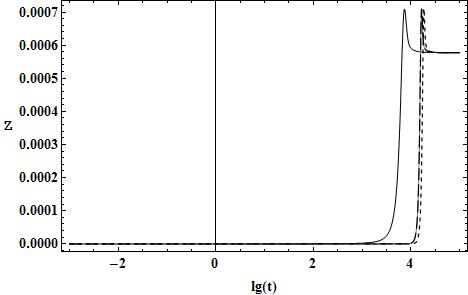}
\caption{The dependency of the potential derivative %
$Z=\dot{\Phi}$ evolution on the fermion scalar charge value. Left to right: $q=0.001;0.1;1;10$. %
Lines for $q=0.1$ and $q=1$ are merged.}
%\label{fig:?}
\end{figure}
\begin{figure}%[tb]
\includegraphics[width=\columnwidth]{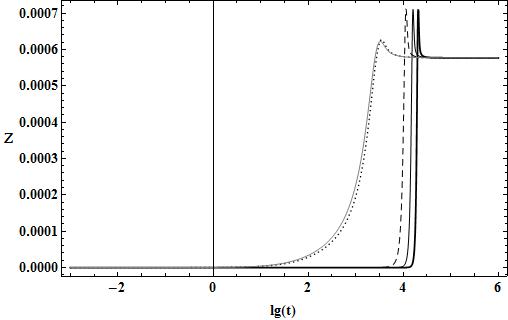}
\caption{The dependency of the potential derivative %
$Z=\dot{\Phi}$ evolution on the scalar charge value for the Boltzmann plasma. Left to right: $q=0.001;0.1;1;10$. Lines for $q=0, m-0.04$ and $q=0.001$ are almost merged.}
%\label{fig:?}
\end{figure}
\begin{figure}%[tb]
\includegraphics[width=\columnwidth]{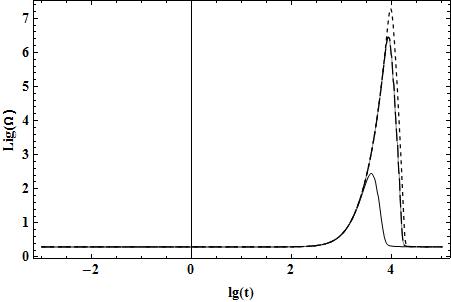}
\caption{The dependency of the invariant cosmological acceleration %
$\Omega$ evolution on the value of the fermion scalar charge. Bottom to top: $q=0.001;0.1;1;10$.}
%\label{fig:?}
\end{figure}
\begin{figure}%[tb]
\includegraphics[width=\columnwidth]{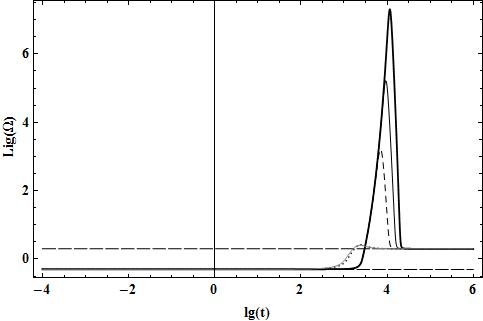}
\caption{The dependency of the invariant cosmological acceleration %
$\Omega$ evolution on the value of the scalar charge for the Boltzmann plasma. Bottom to top: $0.001;0.1;1;10$.}
%\label{fig:?}
\end{figure}
For the Fermi gas lines for $q=0.1$ and $q=1$ are merged: horizontal asymptote corresponds to the $\Omega=+1$. For a Boltzmann gas lines with $q=0, m-0.04$ and $q=0.001$ practically merge. The horizontal dashed lines correspond to the values of $\Omega=-1$ (bottom) and $\Omega=+1$ (top).

\begin{figure}%[tb]
\includegraphics[width=\columnwidth]{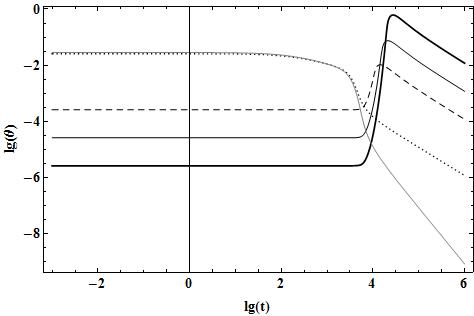}
\caption{The dependency of decimal logarithm of the Boltzmann gas temperature $\theta$ evolution on the value of the scalar particle charge. Bottom to top: $q=0;0.001;0.1;1;10$.}
\label{fig:lgtheta1}
\end{figure}
\begin{figure}%[tb]
\includegraphics[width=\columnwidth]{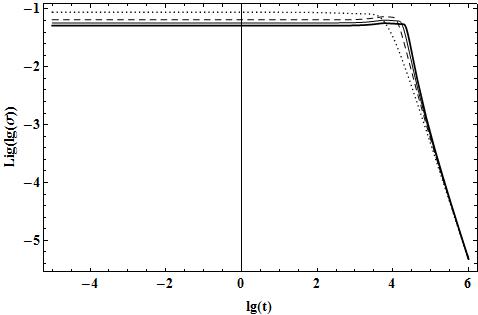}
\caption{The dependency of the scalar charge density function $\mathrm{Lig}\lg(\sigma)$ evolution on the scalar charge value for the Boltzmann gas. Bottom to top: $q=0;0.001;0.1;1;10$.}
\label{fig:lgsigmaB2}
\end{figure}

\subsection{The Comparison with a Case of the Scalar Field. The Relativistic Fermions $p^0_F=3; m=1, m_s=0.1$,  %
$q=0.1$; $\Phi_0=-0.35$}
Let us present the results of the  comparison for the systems with classic and phantom scalar fields. In this case we will take the same initial conditions: (Fig. 15--16). %%__
This case corresponds to the following values of invariants introduced above: %
$\varkappa^0_F=1/3.581315560; \eta^0_S=0.000021697$. In contrast to the above cases, the potential of the scalar %
field is negative (Fig. 23 - 26). It can be seen that classical field in contrast to phantom field, reveals the oscillatory nature. This property of the classical field was found earlier in articles \cite{YuMif,YuMif11}.
\begin{figure}%[tb]
\includegraphics[width=\columnwidth]{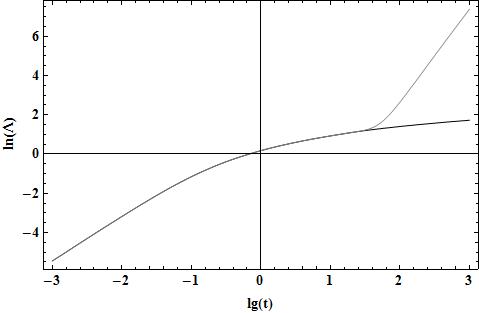}
\caption{Evolution of the scale function $\lg\Lambda(t)$. %
The grey line shows the phantom field and the black line shows the classic field.}
\label{fig:L3}
\end{figure}
\begin{figure}%[tb]
\includegraphics[width=\columnwidth]{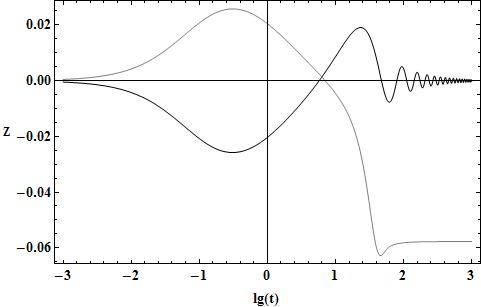}
\caption{Evolution of the potential derivative $Z=\dot{\Phi}$. The grey line shows the phantom field and the black line shows the classic field.}
\label{fig:dF3}
\end{figure}
\begin{figure}[tb]
\includegraphics[width=\columnwidth]{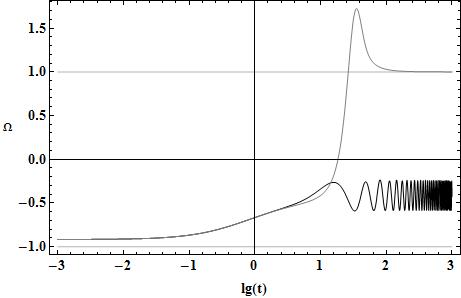}
\caption{Evolution of the invariant cosmological acceleration $\Omega$. The grey line shows the phantom field and the black line shows the classic field.}
\label{fig:O3}
\end{figure}
\begin{figure}[tb]
\includegraphics[width=\columnwidth]{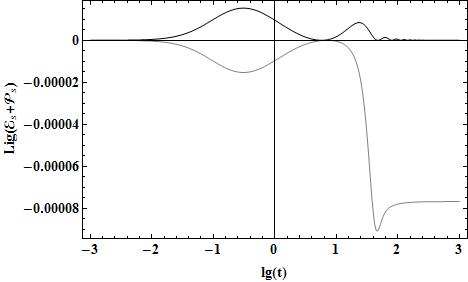}
\caption{Evolution of the invariant total energy density $\mathrm{Lig}\mathcal{E}$. The grey line shows the phantom field and the black line shows the classic field.}
\label{fig:dF3}
\end{figure}

\section{The Conclusion}
Thus, we can state that at large times $\sim 10^5 t_{Pl}$ the cosmological evolution
of matter based on a statistical system of scalar charged particles with a phantom
interaction does not vary from the evolution of matter with a minimal scalar interaction.
However at smaller times $10^5 t_{Pl}$ the evolution of matter with a nonminimal scalar
interaction is characterized by a large number of the behavior types with respect to the
matter with a minimal scalar interaction and also is characterized by the presence of
phantom bursts. In contrast to a system with classical scalar interaction, in the system
of fermions with a phantom scalar interaction the microscopic oscillations with a frequency
of order $m_s$ are not emerged. %
The results of numerical simulation depend insignificantly on the particle statistics: in fact, almost identical results were obtained for one-component degenerate Fermi systems and for two-component Boltzmann plasma. Numerical simulation of the cosmological evolution of the statistical system of scalar charged particles revealed two brand new features of such cosmological model behavior with respect to minimal models. The first difference is in the appearance of giant bursts (up to $10^{10}$) of the invariant cosmological acceleration $\Omega$ at the time interval $8\cdot10^3\div2\cdot10^4 t_{Pl}$. As we noted, the presence of such bursts requires significant modification of quantum theory for generation of cosmological gravitational perturbations. Certainly, this is true if at least a small component of scalar charged particles exists at these stages of the Universe evolution. The second characteristic feature of the cosmological model with a scalar particle interaction is strong heating ($3\div 8$ orders of magnitude) of a statistical system at these times of evolution. Such a strong heating, on the one hand, automatically causes particle-antiparticle pair production, and thereby it generates an initial cosmological plasma. On the other hand, heating requires accounting of the appearance of scalar neutral particles in equilibrium plasma, i.e., requires modifications of  the considered cosmological model. These issues will be considered by us in next paper.

\end{document}